\documentclass[reprint,amsmath,amssymb, aps]{revtex4-1}
\usepackage{graphicx}
\usepackage{dcolumn}
\usepackage{bm}
\usepackage{graphicx}
\usepackage{color}
\usepackage{subfigure}
\usepackage{afterpage}
\usepackage{booktabs}
\usepackage{blkarray}
\usepackage{bm}
\usepackage{times} 
\usepackage{braket}
\usepackage[colorlinks,
citecolor=blue,linkcolor=blue,urlcolor=blue]{hyperref}
\hypersetup{linktocpage}
\usepackage{cancel}
\usepackage{mdframed}
\newcommand{\bea}{\begin{eqnarray}}
\newcommand{\eea}{\end{eqnarray}}

\definecolor{bblue}{rgb}{0.36, 0.54, 0.66}

\newcommand{\be}{\begin{equation}}
\newcommand{\ee}{\end{equation}}

\newcommand{\beA}{\begin{align}}
\newcommand{\eeA}{\end{align}}

\newcommand{\bV}{\begin{pmatrix}}
\newcommand{\eV}{\end{pmatrix}}

\newcommand{\dd}{\mathrm{d}}
\newcommand{\de}{\partial}

\newcommand{\Ha}{\hat{H}} 

\newcommand{\tot}{\text{tot}}

\definecolor{graybox}{rgb}{0.9,0.9,0.9}
\newmdenv[backgroundcolor=graybox,linecolor=graybox]{compl}

\newcommand{\qqq}{\end{document}}

\newcommand{\I}{\textbf{i}}
\begin{document}

\title{Generalized hydrodynamics of the repulsive spin-$\frac{1}{2}$ Fermi gas}
\author{Stefano Scopa$^1$}
\email{sscopa@sissa.it}
\author{Pasquale Calabrese$^{1,2}$}
\author{Lorenzo Piroli$^{3}$}
\affiliation{$^1$~SISSA and INFN, via Bonomea 265, 34136 Trieste, Italy}
\affiliation{$^2$~International Centre for Theoretical Physics (ICTP), I-34151, Trieste, Italy}
\affiliation{$^3$~Philippe Meyer Institute, Physics Department,  \'Ecole Normale Sup\'erieure (ENS),
	Universit\'e PSL, 24 rue Lhomond, F-75231 Paris, France}
\date{\today}
\begin{abstract}
We study non homogeneous quantum quenches in a one-dimensional gas of repulsive spin-$1/2$ fermions, as described by the integrable Yang-Gaudin model. By means of generalized hydrodynamics (GHD), we analyze in detail the real-time evolution following a sudden change of the confining potential. We consider in particular release protocols and trap quenches, including a version of the quantum Newton's cradle. At zero temperature, we employ a simplified phase-space hydrodynamic picture to characterize the dynamics of the particle- and spin-density profiles. Away from zero temperatures, we perform a thorough numerical study of the GHD equations, and provide quantitative predictions for different values of the temperature, external magnetic field, and chemical potential. We highlight the qualitative features arising due to the multi-component nature of the elementary excitations, discussing in particular effects of spin-charge separation and dynamical polarization. 
\end{abstract}

\maketitle
\section{Introduction}\label{intro}

One-dimensional ($1$D) many-body quantum physics is a fascinating subject, characterized by peculiar phenomena and rich mathematical structures~\cite{Giamarchi-book, Tsvelik-book}. Despite its long history, it continues to pose many difficult questions and motivate intense theoretical research. This is especially true out of equilibrium, also due to the recent advances and possibilities introduced by cold-atom physics~\cite{bloch2008,Polkovnikov2011,Guan2013,Guan2022}. In this context, {integrable systems} play an important role, providing simplified models where powerful analytic techniques can be applied.

Recently, a milestone was achieved in the description of integrable systems out of equilibrium, with the introduction of a novel {generalized hydrodynamics} (GHD)~\cite{Bertini2016,Castro-Alvaredo2016}, extending conventional hydrodynamics by taking into account all the local conservation laws associated with integrability. Over the past few years, the theory has proven to be extremely versatile, allowing us to tackle questions beyond the scope of its traditional formulation~(see e.g.,~Refs.~\cite{ghd-rev,ghd-notes,Essler-ghd-rev,ghd-JB,DeNardis-rev,Bulchandani-rev} for recent reviews), including weak-integrability breaking \cite{Bastianello2020b,Bastianello2021}, atom losses \cite{Bouchoule2020}, diffusive corrections to ballistic transport~\cite{DeNardis2018,DeNardis2019,Medenjak2020,Durnin2021}, and large-scale quantum fluctuations~\cite{Ruggiero2020,Collura2020, Scopa2021a,Scopa2022,Scopa2022b, Scopa2022c,Ruggiero2022,Fagotti2017,Fagotti2020}.
Compared to previous analytic approaches to integrable systems, including the quench-action~\cite{caux2013time, caux2016quench} or the quantum-transfer-matrix method~\cite{piroli2017from,Piroli2017}, GHD allows us to study different non-equilibrium protocols breaking translation symmetry, such as {trap quenches} or {trap release}. This makes GHD powerful enough to analyze actual cold-atomic experimental settings, as beautifully demonstrated by the recent papers~\cite{Schemmer2019,Malvania2021}.

So far, GHD has been mostly applied, with a few exceptions~\cite{Nozawa2020,Nozawa2021,Mestyan2019,Scopa2021,Moller-2component}, to integrable models with an elementary Bethe ansatz description, {i.e.}, displaying a single-type of quasiparticle excitation. On the other hand, an interesting class of models features multicomponent quasiparticles, whose analysis requires a more sophisticated \emph{nested} Bethe ansatz approach. A well-known example is the Yang-Gaudin model, describing a one-dimensional ($1$D) gas of repulsive spinful fermions. These systems are typically characterized by a richer phenomenology, as exemplified by the {spin-charge separation} (SCS) effect~\cite{Giamarchi-book,Recati2003,Recati2003b,Liu2005,Lars2005,Bohrdt2018,Boll2016,Yang2018,Vijayan2020,Barfnecht2019}. Furthermore, their study is also obviously timely from the point of view of cold-atom experiments, which allow us, for instance, to realize quantum gases with tunable spin~\cite{pagano2014one} {and interactions \cite{Senaratne2022}}.

The aim of this paper is to conduct a thorough investigation of the GHD description of the prototypical Yang-Gaudin model. We provide quantitative predictions for a number of experimentally relevant nonequilibrium protocols and highlight the qualitative features arising due to the multicomponent nature of the elementary excitations. Our paper expands on previous findings reported in Ref.~\cite{Scopa2021}, which focused on the analysis of finite-temperature SCS effects induced by pulse perturbation.

The paper is organized as follows. In Sec.~\ref{model}, we briefly review the $1$D Yang-Gaudin model and its exact solution based on the nested Bethe ansatz. In Sec.~\ref{sec:review_section}, we review the GHD equations, while in Sec.~\ref{sec:tra_quenches} we apply them to the study of different quench protocols, including a harmonic trap release and a quartic-to-harmonic trap quench. Section~\ref{zeroT} is devoted to the limit of zero temperature, reviewing, in particular, its simplified GHD description. Next, in Sec.~\ref{sec:zero_t_features} we focus on dynamical features of the gas arising due to the multi-component nature of the elementary excitations, which can be analyzed in detail at zero temperature. They include a dynamical polarization and the spin-charge separation effects. Finally, Sec.~\ref{conclusions} contains a summary of our work and our conclusions.

\section{The Yang-Gaudin model}\label{model}
We consider a $1$D system of size $\ell$ containing a gas of spin-$\frac{1}{2}$ particles with repulsive contact interactions, described by the Yang-Gaudin Hamiltonian \cite{Yang1967,Gaudin1967}
\begin{align}
\Ha=&-\int_{0}^{\ell} \dd x \sum_{\sigma=\pm} \hat{\Psi}^\dagger_\sigma(x)\left(\de^2_x +\mu +\sigma h\right)\hat{\Psi}_\sigma(x)\nonumber\\
&+ c\int_{0}^{\ell} \dd x  \ \hat\Psi_{+}^\dagger(x)\hat\Psi_-^\dagger(x)\hat\Psi_+(x)\hat\Psi_-(x),\label{YG}
\end{align}
where $\sigma=\pm$ denote the two spin components, $c>0$ is the interaction coupling, $\mu$ is the chemical potential and $h$ is the external magnetic field . The fields $\hat\Psi^\dagger_\sigma$, $\hat\Psi_\sigma$ are creation and annihilation fermionic operators satisfying canonical anticommutation relations
\be
\{\hat\Psi^\dagger_\sigma(x),\hat\Psi_{\sigma^\prime}(x^\prime)\}=\delta_{\sigma,\sigma^\prime}\delta(x-x^\prime).
\ee
It is well-known that the Yang-Gaudin model can be solved by the Bethe ansatz \cite{Yang1967,Gaudin1967,Takahashi1999,Batchelor2010,Guan2013}. Focusing on the sector with $N$ fermions, $M$ of which have spin down, the Hamiltonian \eqref{YG} can be written in the first-quantized form 
\be\label{eigenproblem}
\hat{\cal H}_{N,M}= -\sum_{j=1}^N \left(\de^2_{x_j} +\mu\right) - h(N-2M) +2c \sum_{i<j} \delta(x_i-x_j)
\ee
with associated many-body eigenstates
\be
\hat{\cal H}_{N,M} \ \psi_{\bm{k},\bm\lambda}(\vec{x}) = E({\bm{k},\bm\lambda})\ \psi_{\bm{k},\bm\lambda}(\vec{x}),
\ee
$\vec{x}=(x_1,\dots,x_N)$, whose explicit form can be found e.g.~in Ref.~\cite{Yang1967}. Importantly, these eigenstates are labeled by two sets of spectral parameters (or rapidities), namely $\bm{k}$ and $\bm{\lambda}$, which identify two different quasiparticles species. The first set $\bm{k}=\{k_1,\dots,k_N\}$ is composed by the quasimomenta of the physical particles in the system while the second set $\bm\lambda=\{\lambda_1,\dots,\lambda_M\}$ is related to the spin degrees of freedom. Imposing periodic boundary conditions on the system, from Eq.~\eqref{eigenproblem} one can derive the following algebraic Bethe equations 
\begin{subequations}\label{eq:bethe_equations}
\be
e^{\I k_j \ell}= \prod_{\alpha=1}^M \frac{k_j -\lambda_\alpha+\I c/2}{k_j -\lambda_\alpha-\I c/2}\,,
\ee
\be
\prod_{j=1}^N \frac{\lambda_\alpha-k_j+\I c/2}{\lambda_\alpha-k_j -\I c/2} = \prod_{\beta\neq \alpha,\beta=1}^M \frac{\lambda_\alpha-\lambda_\beta+\I c}{\lambda_\alpha-\lambda_\beta-\I c}\,,
\ee
\end{subequations}
which constrain the quantum numbers $\bm{k},\bm\lambda$ to take only some specific values. For the repulsive gas ($c>0$), $k_j$ are real numbers while, for $\ell\gg 1$, the rapidities $\lambda_\alpha$ are grouped together in symmetric patterns around the real axis called strings \cite{Takahashi1999}, which correspond to bound states of spin quasiparticles. For instance, a string of size $n$ is composed by the rapidities
\be\label{string}
\lambda_{\alpha,j}= \lambda_\alpha^n + \I(n+1-2j)c/2, \quad j=1,\dots, n\,,
\ee
with $\lambda_\alpha^n\in\mathbb{R}$ known as string center.

In the thermodynamic limit where $\ell,N,M\to \infty$ at fixed density $N/\ell$ and $M/\ell$, the spectrum of the model becomes densely populated and the rapidities $\bm{k}$, $\bm\lambda$ can be replaced by the density distributions (or root densities)
\be
\rho_1(k_j)\propto\frac{1}{\ell(k_{j+1}-k_j)} ,  \ \rho_{2,n}(\lambda_{\alpha}^n)\propto\frac{1}{\ell (\lambda_{\alpha+1}^n -\lambda_\alpha^n)}.
\ee
{The root density $\rho_{2,n}$ for any $n\geq 1$ depends only on the value of the string center $\lambda\equiv \lambda^n_\alpha$.}
Taking the logarithm of the Bethe equations~\eqref{eq:bethe_equations}, and the thermodynamic limit, we arrive at the Bethe-Gaudin-Takahashi (BGT) equations
\begin{subequations}\label{BGT}
	\be\label{BGT1}
	\rho_1^\text{tot}(k)= \frac{1}{2\pi} + \sum_{n=1}^\infty[\phi_n\ast \rho_{2,n}](k)\,,
	\ee
	\be\label{BGT2}
	\rho_{2,n}^\text{tot}(k) =[\phi_n\ast\rho_1](k) - \sum_{m=1}^\infty [\Phi_{n,m}\ast \rho_{2,m}](k).
	\ee
\end{subequations}
Here, we defined 
\be
\phi_n(k)=\frac{1}{\pi} \ \frac{2nc}{(nc)^2 + 4k^2},
\ee
\be\begin{split}
	\Phi_{n,m}(k)&=(1-\delta_{n,m})\phi_{|n-m|}(k) + 2\phi_{|n-m|+2}(k)+\dots\\
	& + 2\phi_{n+m-2}(k) + \phi_{n+m}(k)\,,
\end{split}\ee
and used the notation 
\be
[g_1\ast g_2](k)=\int_{-\infty}^\infty \dd k' \ g_1(k-k') \ g_2(k').
\ee
Furthermore, we have introduced the total densities $\rho_1^\text{tot}\equiv \rho_1+\rho^h_1$, $\rho_{2,n}^\text{tot}\equiv \rho_{2,n}+\rho^h_{2,n}$, where $\rho_1^h$ and $\rho_{2,n}^h$ are the densities of unoccupied rapidities, or {\it holes}. 

Equations~\eqref{BGT1} and \eqref{BGT2} do not uniquely determine the distributions $\rho_1, \rho_{2,n}$ and an additional set of equations must be derived, depending on the macroscopic state of the model. Let us consider, for instance, the case where the system is at thermal equilibrium and at temperature $T$. Introducing {the free-energy functional \cite{Takahashi1999}
\be\begin{split}
&{\cal G}(T,\mu,h)/\ell\equiv\int_{-\infty}^\infty \dd k \big\{ (k^2-\mu-h) \rho(k)- T s_1(k)\big\}\\
&+ \sum_{n=1}^\infty \int_{-\infty}^\infty \dd \lambda\big\{2nh\ \rho_{2,n}(\lambda)-T  s_{2,n}(\lambda)\big\}
\end{split}\ee
with entropy densities
\be\begin{split}
&s_1= \rho_1\log(1+ e^{\epsilon_1/T}) + \rho^h_1\log(1+e^{-\epsilon_1/T});\\[3pt]
&s_{2,n}= \rho_{2,n}\log(1+ e^{\epsilon_{2,n}/T}) + \rho^h_{2,n}\log(1+e^{-\epsilon_{2,n}/T}),
\end{split}\ee
and dressed energies
\be\label{ratio}
\frac{\rho_1^h}{\rho_1}= \exp(\epsilon_1(k)/T), \quad \frac{\rho^h_{2,n}}{\rho_{2,n}}=\exp(\epsilon_{2,n}(\lambda)/T)\,,
\ee
one can derive the following thermodynamic Bethe ansatz (TBA) equations {by minimizing ${\cal G}$ according to Eqs.~\eqref{BGT}} \cite{Takahashi1999,Takahashi1971,Takahashi1973}
\begin{subequations}\label{TBA}
\be\label{TBA1}
\frac{\epsilon_1(k)}{T}= \frac{k^2-\mu-h}{T} -\sum_{n=1}^\infty [\phi_n \ast \log(1+e^{-\epsilon_{2,n}/T})](k),
\ee
\be\label{TBA2}\begin{split}
\frac{\epsilon_{2,n}(k)}{T}=& \frac{2nh}{T} -[\phi_n \ast \log(1+e^{-\epsilon_1/T})](k) \\
&+ \sum_{m=1}^\infty [\Phi_{n,m}\ast \log(1+e^{-\epsilon_{2,m}/T})](k).
\end{split}\ee
\end{subequations}

\begin{figure}[t!]
\centering
(a)\\
\includegraphics[width=0.375\textwidth]{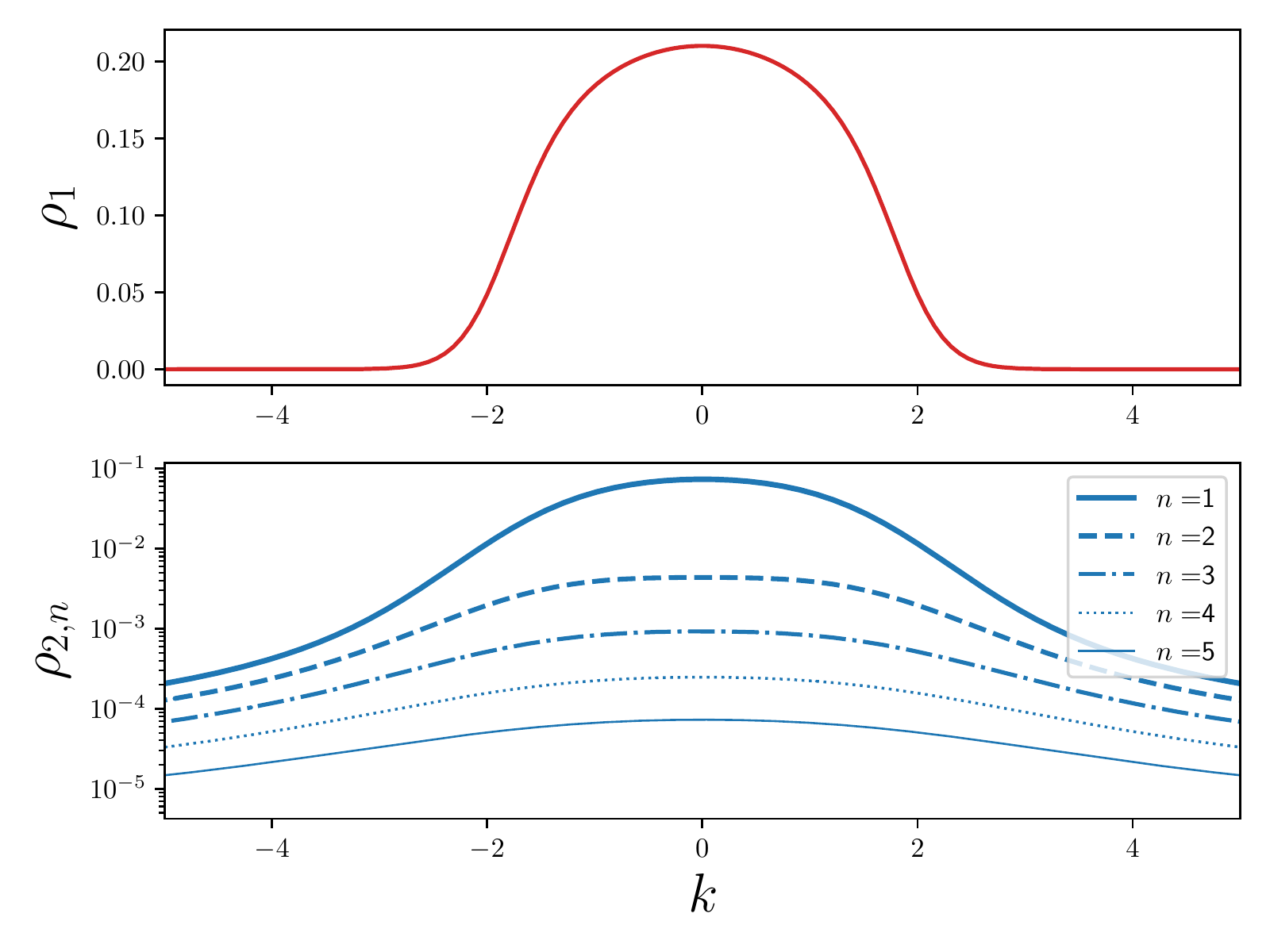}\\
(b)\\
\includegraphics[width=0.375\textwidth]{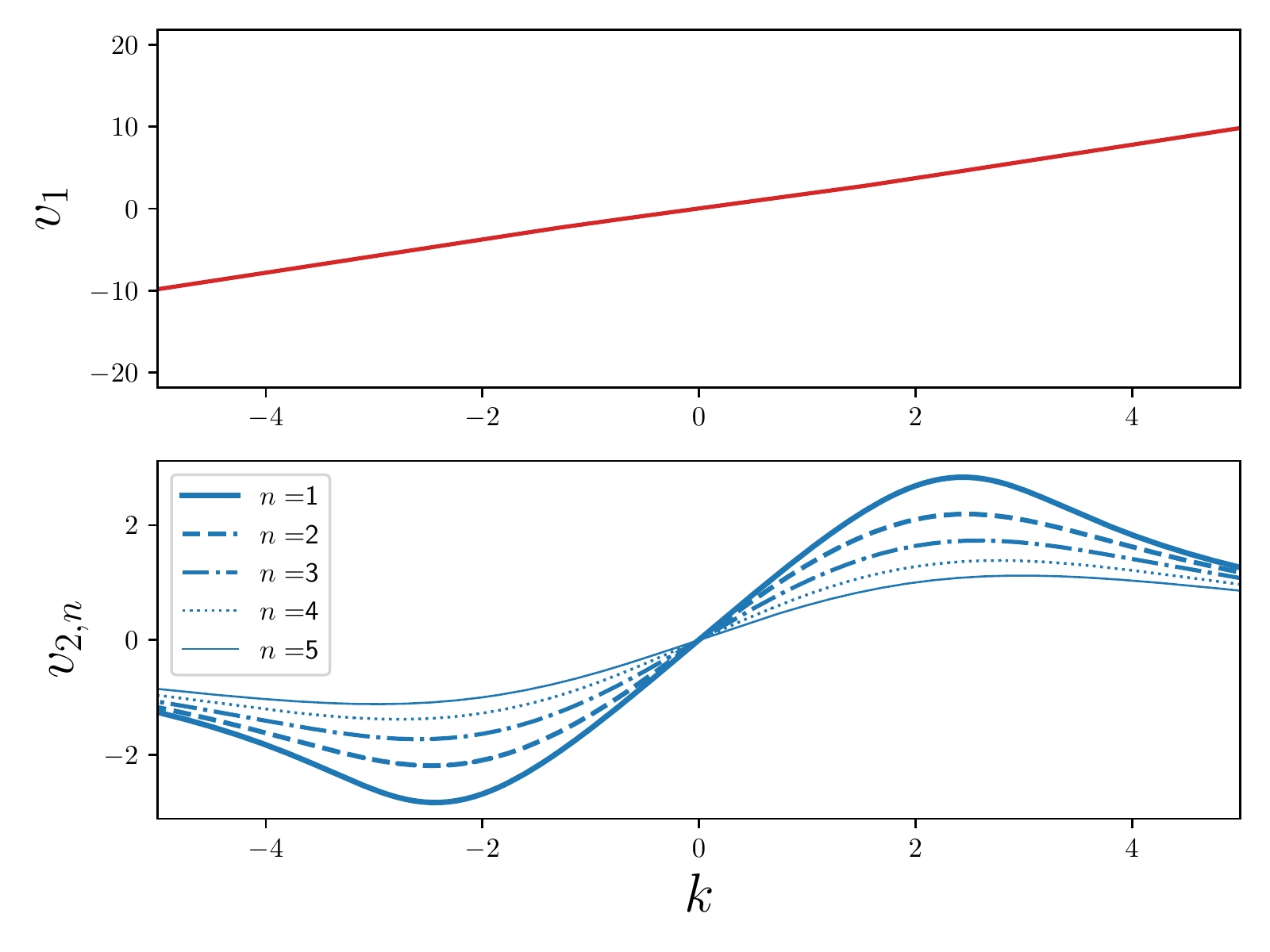}
\caption{\label{fig:root-dens} (a) Root densities $\{\rho_1, \rho_{2,n}\}$ and (b) effective velocities $\{v_1, v_{2,n}\}$ as functions of the rapidity $k$, obtained from the numerical solution of Eqs.~\eqref{BGT} (panel a) and Eqs.~\eqref{BEL} (panel b). The curves are obtained setting $c=1$, $\mu=2$, $h=0.5$ and $T=1$.}
\end{figure}
For later convenience, we also define the effective velocities of the two quasiparticles species as \cite{Bonnes2014,Bertini2016,Castro-Alvaredo2016,Mestyan2019}
\be\label{Vel}
v_1(k)=\frac{\de_k\epsilon_1(k)}{2\pi \rho^{\rm tot}_1(k)}, \quad v_{2,n}(\lambda)=\frac{\de_k\epsilon_{2,n}(\lambda)}{2\pi \rho_{2,n}^{\rm tot}(\lambda)},
\ee
which can be obtained as the solution to the equations \cite{Bonnes2014}
\begin{subequations}\label{BEL}
\be\label{vel1}
v_1(k) \rho^{\rm tot}_1(k)= \frac{k}{\pi} + \sum_{n=1}^\infty [\phi_n\ast v_{2,n} \rho_{2,n}](k);
\ee
\be\label{vel2}\begin{split}
v_{2,n}(k) \rho^{\rm tot}_{2,n}(k)&=[\phi_n\ast v_1 \rho_1](k) \\
& -\sum_{m=1}^\infty [\Phi_{n,m}\ast v_{2,m} \rho_{2,m}](k).
\end{split}\ee
\end{subequations}
Equations~\eqref{BGT},~\eqref{TBA} and \eqref{BEL} can be solved numerically with standard iterative methods. We report an explicit example of their solution in Fig.~\ref{fig:root-dens}.\\

Finally, let us consider an arbitrary function of the rapidities $\vec\zeta=\{\zeta_1(k), \zeta_{2,n}(\lambda)\}$. Introducing the \emph{filling functions}
\be
\theta_1\equiv \frac{\rho_1}{\rho_1^\tot}=\frac{1}{1+e^{\epsilon_1/T}}; \  
\theta_{2,n}\equiv \frac{\rho_{2,n}}{\rho_{2,n}^\tot}=\frac{1}{1+e^{\epsilon_{2,n}/T}}\,,
\ee 
we can define the dressed function $\vec \zeta^{\rm dr}$ as the solution to the equations
\begin{subequations}\label{dressing}
\be
\zeta_1^\text{dr}(k)=\zeta_1(k) + \sum_{n=1}^\infty [\phi_n \ast \zeta^\text{dr}_{2,n} \theta_{2,n}](k),
\ee
\be
\zeta^\text{dr}_{2,n}(k)=\zeta_{2,n}(k) +[\phi_n\ast \zeta^\text{dr}_1\theta_{1}](k)-  \sum_{m=1}^\infty [\Phi_{n,m} \ast \zeta^\text{dr}_{2,m}\theta_{2,m}](k).
\ee
\end{subequations}
From these definitions, one can easily recover the usual relations for the derivative of the dressed momentum
\be\label{deriv-mom}
(\de_k\vec{p})^\text{dr}=[ \{1, 0\}]^\text{dr}\equiv 2\pi\vec{\rho}^\text{\ tot}\,,
\ee
and of the derivative of the dressed energy $e(k)=\{k^2-\mu-h, 2n h\}$
\be
(\de_k\vec{e})^\text{dr}=[ \{k, 0\}]^\text{dr}\equiv \de_k \vec{\epsilon}\,,
\ee
such that the effective velocities in Eq.~\eqref{Vel} are $\vec{v}=\de_k \vec{\epsilon}/(\de_k\vec{p})^\text{dr}$ (in this notation, the ratio of vector quantities is intended as the ratio of each component).\\

Given the set of root densities $\vec\rho(k)=\{\rho_1(k),\rho_{2,n}(k)\}$ and of effective velocities $\vec{v}(k)=\{v_1(k),v_{2,n}(k)\}$, it is possible to compute the conserved charges and current densities. More precisely, for any conserved charge $\hat{\cal Q}$ one can write \cite{Castro-Alvaredo2016,Bertini2016,Mestyan2019}
\begin{align}
q=\frac{\braket{\hat{\cal Q}}}{\ell}&=\int_{-\infty}^\infty \dd k \ {\tt q}_1(k) \rho_1(k) \nonumber\\
&+ \sum_{n=1}^\infty \int_{-\infty}^\infty \dd k  \ {\tt q}_{2,n}(k) \rho_{2,n}(k)\,,
\label{charge}
\end{align}
and the current \cite{Borsi2019,Pozsgay2020, Pozsgay2020b,Borsi2021, Bajnok2020, Vu2019,Yoshimura2020,Urichuk2019}
\be\label{curr}\begin{split}
j_q=&\int_{-\infty}^\infty \dd k \ {\tt q}_1(k) v_1(k) \rho_1(k) \\
&+ \sum_{n=1}^\infty \int_{-\infty}^\infty \dd k  \ {\tt q}_{2,n}(k) v_{2,n}(k) \rho_{2,n}(k)\,,
\end{split}\ee
where $\vec{\tt q}=\{{\tt q}_1, {\tt q}_{2,n}\}$ are the single-particle eigenvalues associated with the charge $\hat{\cal Q}$. For instance, the {particle density and current are given by 
\begin{subequations}
\be
\varrho=\frac{N}{\ell}=\int_{-\infty}^\infty \dd k \ \rho_1(k)
\ee
\be
j_\varrho=\int_{-\infty}^\infty \dd k \ v_1(k)\rho_1(k),
\ee
\end{subequations}
while magnetization density and current are
\begin{subequations}
\be
m=\frac{N-2M}{2\ell}=\varrho/2-\sum_{n=1}^\infty n \int_{-\infty}^\infty \dd k \ \rho_{2,n}(k), 
\ee
\be
j_m=j_\varrho/2-\sum_{n=1}^\infty n \int_{-\infty}^\infty \dd k \ \rho_{2,n}(k) v_{2,n}(k)\,,
\ee
\end{subequations}
{i.e.}, they correspond to $\vec{\tt q}_\varrho=\{1,0\}$ and $\vec{\tt q}_m=\{1/2,-n\}$ respectively. Similarly, other sets of single-particle eigenvalues identify other conserved quantities, e.g.~$\vec{\tt q}_e=\big\{ k^2-\mu-h,  2nh \big\}$ identifies the energy density and its current.}
\begin{figure}[t!]
\centering
\includegraphics[width=0.375\textwidth]{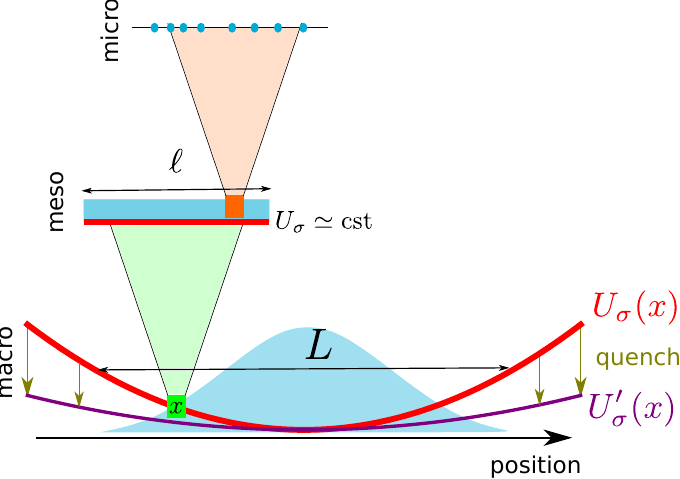}
\caption{\label{fig:disegno}Illustration of the quench protocols considered in this paper -- the spinful gas is coupled to an inhomogeneous external potential $U_\sigma$ that is assumed to be slowly varying at mesoscopic scales $a\ll \ell\ll L$. Under this working hypothesis, the inhomogeneous model at $t<0$ is solved within each fluid cell $x$ by nested Bethe ansatz with local value of potential $U_\sigma(x)\simeq\text{cst}$ (see Sec.~\ref{model}). At $t=0$ we suddenly vary the potential $U_\sigma(x)\overset{t=0}{\to}U'_\sigma(x)$. Requiring the same smoothness hypothesis on the post-quench potential, the problem can be studied in the hydrodynamic limit.}\end{figure}

\section {The Quench setup and  the GHD equations}
\label{sec:review_section}
\subsection{The quench protocols}\label{quench}
Let us discuss the class of quench protocols of interest in  this paper. First, we prepare the system in an equilibrium state of the Yang-Gaudin Hamiltonian in the presence of a nonhomogeneous potential $U_\sigma(x)$ confining the gas in the spatial region $-\frac{L}{2}\leq x \leq \frac{L}{2}$, namely
\be\begin{split}\label{H}
\Ha=&\int_{-\frac{L}{2}}^{\frac{L}{2}} \dd x \sum_{\sigma=\pm} \hat{\Psi}^\dagger_\sigma(x)\left[-\de^2_x + U_\sigma(x)\right]\hat{\Psi}_\sigma(x)\\
&+ c\int_{-\frac{L}{2}}^{\frac{L}{2}} \dd x  \ \hat\Psi_{+}^\dagger(x)\hat\Psi_-^\dagger(x)\hat\Psi_+(x)\hat\Psi_-(x),\end{split}\ee
where the external potential reads
\be
U_\sigma(x)=V(x)-\mu-\sigma (h+g(x)).
\ee
The presence of a trap potential $V(x)$ and of an inhomogeneous magnetic field $g(x)$ spoils the exact solvability of the model discussed in Sec.~\ref{model}.  However, by assuming that the potential $U_\sigma(x)$ is a slowly-varying function on macroscopically large scales, it is possible to focus on a mesoscopic description of the model over fluid cells of size $\ell$ such that
\be\label{sep-of-scales}
a\ll \ell \ll L\,,
\ee
where $a$ is the microscopic scale of the problem, typically of order of the inverse local density of particles $a\sim {\cal O}(\varrho(x)^{-1})$ in the fluid cell labeled by $x$. Under such a scale separation hypothesis, at each fluid cell $x$ the system appears locally homogeneous, \emph{i.e.}, $U_\sigma(x)\simeq \text{cst}$,  but it still contains a sufficiently large number of constituents. As a result, one obtains a thermodynamic description of the intial non-homogeneous state in terms of some local root densities $\vec\rho(x,k)$.\\

 Next, we consider the quench protocol where the shape of the initial potential is suddenly changed at time $t=0$ 
\be
U_\sigma(x)\; \overset{t=0}{\mapsto} \; U'_\sigma(x)\equiv V'(x)-\mu'+\sigma(h'+g'(x))\,,
\ee
and we investigate the non-equilibrium dynamics generated by the Hamiltonian \eqref{H} with potential $U'_\sigma(x)$. Note that prime denotes the post-quench fields, not derivatives. Importantly, we require the same smoothness assumptions \eqref{sep-of-scales} for the post-quench potential $U'_\sigma(x)$ so that our quench problem is suitably described by the Euler hydrodynamic equations detailed below. An illustration of our working assumptions and of the quench protocols is shown in Fig.~\ref{fig:disegno}.

\subsection{The GHD equations}\label{sec:GHD}
In this section, we derive the set of GHD equations for the evolution of the root densities $\vec\rho=\{\rho_1,\rho_{2,n}\}$ in the presence of a non-homogeneous potential $U'_\sigma(x)$. They take the form 
\be\label{GHD}
\begin{split}
&\de_t \rho_1 + \de_x (\nu_1 \rho_1) + \de_k(a_1 \rho_1) =0\ , \\[4pt]
&\de_t \rho_{2,n} + \de_x (\nu_{2,n} \rho_{2,n}) + \de_k(a_{2,n} \rho_{2,n})=0,
\end{split}\ee
or equivalently in terms of the filling functions $\vec\theta=\{\theta_1,\theta_{2,n}\}$
\be\label{ghd}
\begin{split}
&\left(\de_t  + \nu_1 \de_x  + a_1\de_k\right) \theta_1=0\ ,\\[4pt]
&\left(\de_t  + \nu_{2,n} \de_x  + a_{2,n}\de_k\right)\theta_{2,n} =0,
\end{split}
\ee
with effective velocities $\vec{\nu}=\{\nu_1,\nu_{2,n}\}$ and effective accelerations $\vec{a}=\{a_1,a_{2,n}\}$ to be determined. To this end, we rewrite the post-quench inhomogeneous Hamiltonian \eqref{H} as
\be\begin{split}
\Ha=& \Ha_0 + \int_{-\frac{L}{2}}^{\frac{L}{2}} \dd x \left( V'(x) \hat{N}(x) - g'(x) \hat{M}(x)\right)\,,
\end{split}\ee
where $\Ha_0$ is the homogeneous part of $\Ha$ (obtained setting $V'=g'=0$ in Eq.~\eqref{H}) and 
\be\begin{split}
&\hat{N}(x)\equiv \sum_{\sigma=\pm}\hat\Psi_\sigma^\dagger(x)\hat\Psi_\sigma(x)\,,\\[4pt]
&\hat{M}(x)\equiv \hat\Psi_-^\dagger(x)\hat\Psi_-(x)\,,
\end{split}\ee
 are the total particle-number and the spin-down particle-number operators, respectively.\\

Following Ref.~\cite{Doyon-IGHD}, we introduce the pseudopotential
\be\label{pseudo}
\vec{w}(x,k) \! \equiv  \! \bV \left[V'(x)-g'(x)\right] {\tt q}^{(N)}_1(k) \\[4pt]  2g'(x) {\tt q}^{(M)}_{2,n}(k)\eV \!= \!\bV V'(x)-g'(x)  \\[4pt]  2n g'(x)\eV\,.
\ee
Equation~\eqref{pseudo} follows from the fact that the single-particle eigenvalues associated to the pertubations induced by $\hat{N}$ and $\hat{M}$ are $\vec{\tt q}^{(N)}=\{1,0\}$ and $\vec{\tt q}^{(M)}=\{0,n\}$, respectively~\cite{Doyon-IGHD}. The effective velocities for the model are then obtained from
\begin{align}
\vec{\nu}(x,k)&=\frac{\left(\de_k [\vec{e}(x,k) + \vec{w}(x)]\right)^\text{dr}}{\left(\de_k\vec{p}(x,k)\right)^\text{dr}}\nonumber\\
&=\frac{\de_k \vec{\varepsilon}(x,k) }{2\pi\vec{\rho}^\text{\ tot}(x,k)}\equiv \vec{v}(x,k)\,,
\end{align}
{i.e.}, the effective velocities in Eq.~\eqref{Vel} are not modified by the presence of $U'_\sigma(x)$ since the resulting pseudopotential \eqref{pseudo} does not depend on the rapidity \cite{Doyon-IGHD}. For each fluid cell $x$, the effective velocities $\vec{v}(x,k)$ can be determined by solving Eqs.~\eqref{BEL} with effective chemical potential $\tilde\mu=\mu'-V'(x)$ and magnetic field $\tilde{h}=h'+g'(x)$.\\

The effective accelerations are instead nontrivial and read \cite{Doyon-IGHD}
\be\label{acc-gen}
\vec{a}(x,k)=\frac{\left(-\de_x\vec{w}(x)\right)^\text{dr}}{\left(\de_k \vec{p}\right)^\text{dr}}=\frac{\left(\{ -\de_x V' +\de_x g' , 2n\de_x g'\}\right)^\text{dr}}{2\pi\vec{\rho}^\text{\ tot}(x,k)}.
\ee
This expression greatly simplifies in the presence of a constant magnetic field $g'=0$
\be\label{acc}
\vec{a}(x,k)\equiv \vec{a}(x)= -\de_x V'\ \{1, 1\}\,,
\ee
where we used Eq.~\eqref{deriv-mom}. Note that the trap potential $V'(x)$ induces a forcing term $a_{2,n}=-\de_x V'$ appearing in the spin species even in the absence of inhomogeneous magnetic fields ($g'=0$), due to backreaction effects in the nested dressing operation \eqref{dressing}.\\

The GHD equations~\eqref{ghd} are formally solved with the method of characteristics
\be
\vec\theta(t,x,k)= \vec\theta(0, \tilde{x}(t), \tilde{k}(t))\,,
\ee
with trajectories given by
\be\begin{split}\label{x-tilde}
&\tilde{x}(t)= x -\int_0^t \dd s \ \vec{v}(s,\tilde{x}(s),\tilde{k}(s)),
\\
&\tilde{k}(t)= k -\int_0^t \dd s \ \vec{a}(\tilde{x}(s),\tilde{k}(s)). 
\end{split}\ee
This formal solution is typically used as the starting point for the development of efficient numerical solution of the GHD equations \cite{Bastianello2019, Moller2020}.\\
\section{Trap quenches at finite temperature}\label{sec:tra_quenches}
In this section, we exploit the GHD framework discussed in Sec.~\ref{sec:GHD} to investigate two prototypical classes of quench protocols:\begin{itemize}
\item[{\it 1.}]~{\it release protocols}, where the post-quench dynamics is given by the homogeneous Yang-Gaudin Hamiltonian ($\de_xU'_\sigma =0$);
\item[{\it 2.}]~{\it trap quench protocols}, where the post-quench potential $U'_\sigma(x)=V'(x)-\mu'+\sigma h'$ does not contain inhomogeneous magnetic fields and the effective acceleration are written as in Eq.~\eqref{acc}.
\end{itemize}
\begin{figure}[t!]
\centering
\includegraphics[width=.49\textwidth]{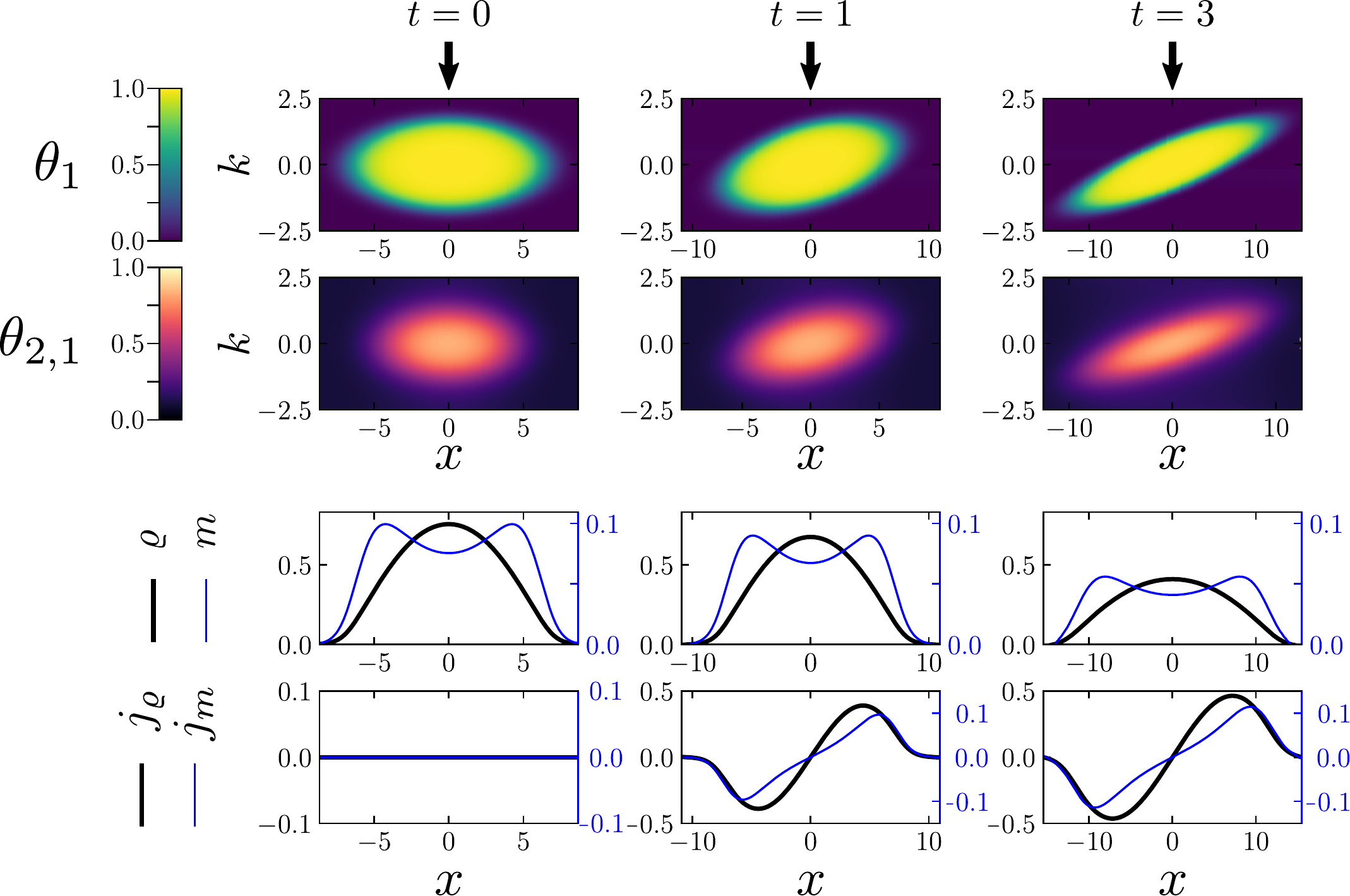}
\caption{\label{fig:GHD-ex1}Solution to the GHD equations \eqref{ghd} for the harmonic trap release: $\omega=0.25\overset{t=0}{\to}0$ at fixed $\mu=2$, $h=0.5$, $c=1$ and temperature $T=0.5$. The first two rows from the top show the time evolution of the filling functions $\theta_1$ and $\theta_{2,1}$ in the rapidity-position plane with time increasing from the leftmost to the rightmost panel. The third and the fourth rows show the evolution of the particles ({\it thick  line}) and magnetization ({\it thin blue line}) density and current as function of the position $x$, with time increasing from the leftmost to the rightmost panel. For a better visualization of the curves, in each panel we plot the two quantities in different scales (left:~particles density; right:~magnetization, {\it in blue}).}
\end{figure}
For the sake of concreteness, we consider an initial harmonic confinement $U_\sigma=\omega^2 x^2-\mu+\sigma h$ and act on the trap frequency $\omega$ at $t=0$, keeping the other parameters constants. First, we study the trap release where $\omega \overset{t=0}{\to} 0$ and we report our results in Fig.~\ref{fig:GHD-ex1}. In this figure, the first two rows from the top show the dynamics of the filling functions in the rapidity-position plane, with time increasing from the leftmost to the rightmost panel. Due to the presence of the confinement, we see that the initial filling functions display an inhomogeneous profile, characterized by a larger occupation of modes around $x=0$ and by small occupations near the edges of the trap where the effective potential barrier $V(x)-\mu$ is higher. This gives rise to the ellipse shape for $\vec\theta(x,k)$ that is shown in the top left panels of Fig.~\ref{fig:GHD-ex1}. We then evolve the filling functions at $t>0$ according to the GHD equations \eqref{ghd}, which in this case consist of a horizontal shift of each quasiparticle of species $a$ by the infinitesimal amount $\dd x\approx v_a(x,k) \dd t$, cf.~Eq.~\eqref{x-tilde}. It is easy to see that such horizontal deformation of the filling functions towards larger values of $x$ is physically associated to the free expansion of the gas after the trap release.  From the GHD evolution of the filling functions, one can determine the root densities via Eqs.~\eqref{BGT} and then the profiles of charges densities and currents using Eqs.~\eqref{charge} and \eqref{curr}. The profiles of particles and magnetization are shown in the third and fourth columns of Fig.~\ref{fig:GHD-ex1}. As one can see,  transport from the center to the edges of the system manifests itself in the flattening of the charge profiles and in the corresponding onset of non-zero currents.\\
\begin{figure}[t!]
\centering
\includegraphics[width=0.475\textwidth]{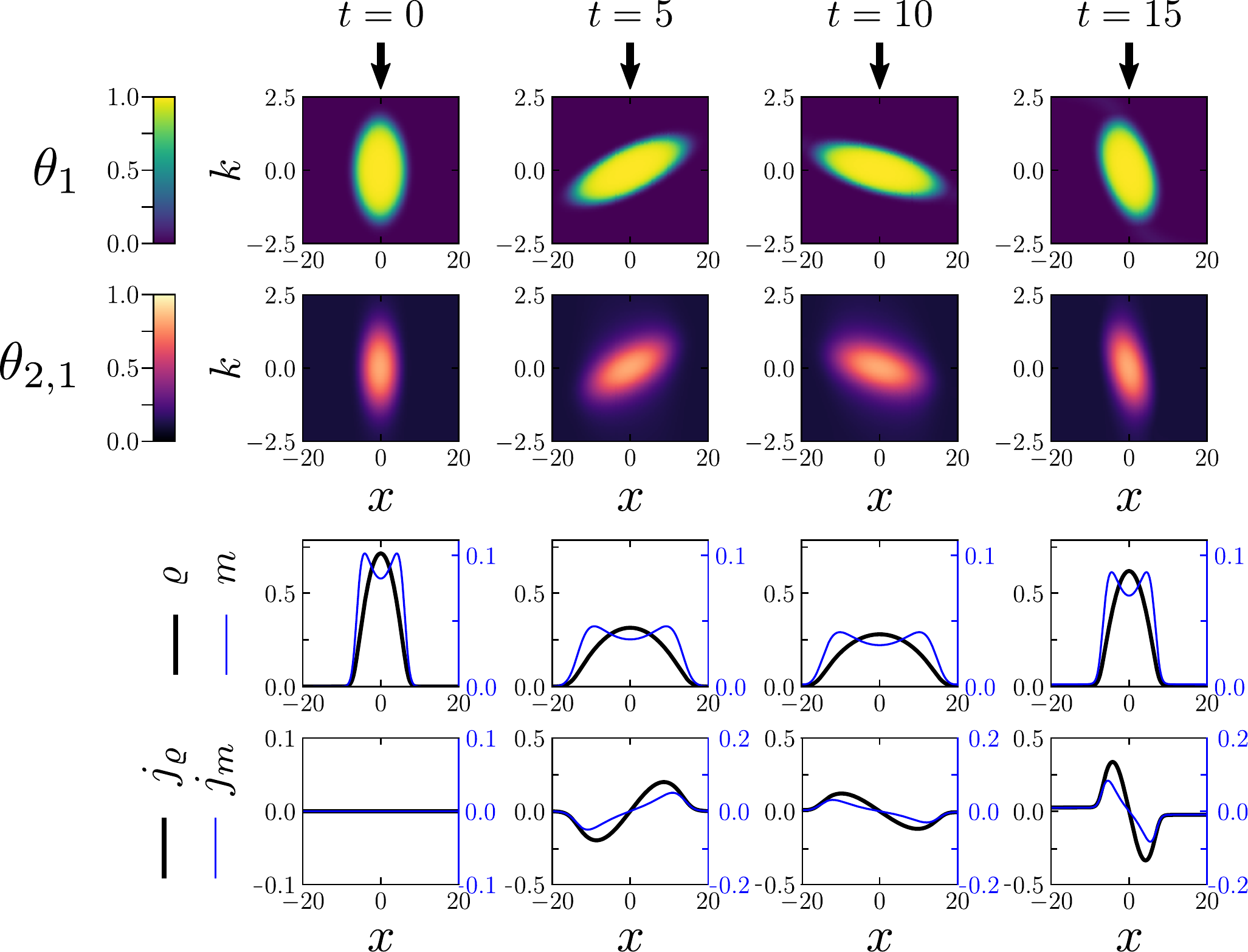}
\caption{\label{fig:GHD-ex2}GHD evolution \eqref{ghd} for the harmonic trap quench: $\omega=0.25\overset{t=0}{\to}\omega'=0.1$ at fixed $\mu=2$, $h=0.5$, $c=1$ and temperature $T=0.5$. We display the results with the same legend of Fig.~\ref{fig:GHD-ex1}. Notice that the presence of a post-quench harmonic trap induces a forcing term $\vec{a}\equiv -2\omega^{\prime \ 2} x\times\{1,1\}$ which leads to the rotation of the filling functions $\vec\theta$ in the rapidity-position plane.}
\end{figure}

Next, in Fig.~\ref{fig:GHD-ex2} we study the case of a harmonic trap quench, where now the trap frequency is suddenly changed to a non-zero value: $\omega \overset{t=0}{\to} \omega'$, with $0<\omega'<\omega$. We initially observe a quasi-free expansion of the gas that resembles the case of Fig.~\ref{fig:GHD-ex1}. After this stage, we see the reflection of the quasiparticles against the edges of the trap and the subsequent recombination of a cloud in the middle of the system. This breathing dynamics is quantitatively captured by the GHD equations \eqref{ghd}, which are now characterized by the presence of a forcing term for the rapidities generating the rotation of $\vec\theta(x,k)$ in the rapidity-position plane.  In the absence of interactions, the period of the motion is $2\pi/\omega'$ but the presence of finite interactions induces a dephasing mechanisms \cite{Caux2018}. The behavior of the charge densities is characterized by contraction and expansion stages and the currents change sign approximatively every half period accordingly, see Fig.~\ref{fig:GHD-ex2}.\\

Lastly, we report the time evolution of the spinful gas during  a quartic-to-quadratic trap quench. In this setting, the system is prepared at equilibrium in a double-well potential such that the initial state is made of two separated clouds, {see Fig.~\ref{fig:GHD-QNC}(a) top left panels. By releasing the quartic potential into the harmonic trap at $t=0$, the two clouds acquire opposite nonzero momenta. Therefore, for $t>0$, they begin to move towards each other, interact and eventually separate again giving rise to a periodic motion.} This realizes a version of the quantum Newton’s cradle, observed in cold-atomic bosonic gases~\cite{Kinoshita2006}. The observed lack of thermalization during the dynamics is rooted in the integrability of the Yang-Gaudin Hamiltonian and reflects in  the quasi-periodic patterns of the charges profiles, see Fig.~\ref{fig:GHD-QNC}(b). Notice that the interactions among particles generate a many-body dephasing that gradually spoils the periodicity of the  motion~\cite{Caux2018}. In Fig.~\ref{fig:GHD-QNC}, the onset of this relaxation mechanism becomes visible at $t\gtrsim 20$, particularly in the filling functions of the spin species. 
\begin{figure*}[t!]
\centering
(a)\\[4pt]
\includegraphics[width=\textwidth]{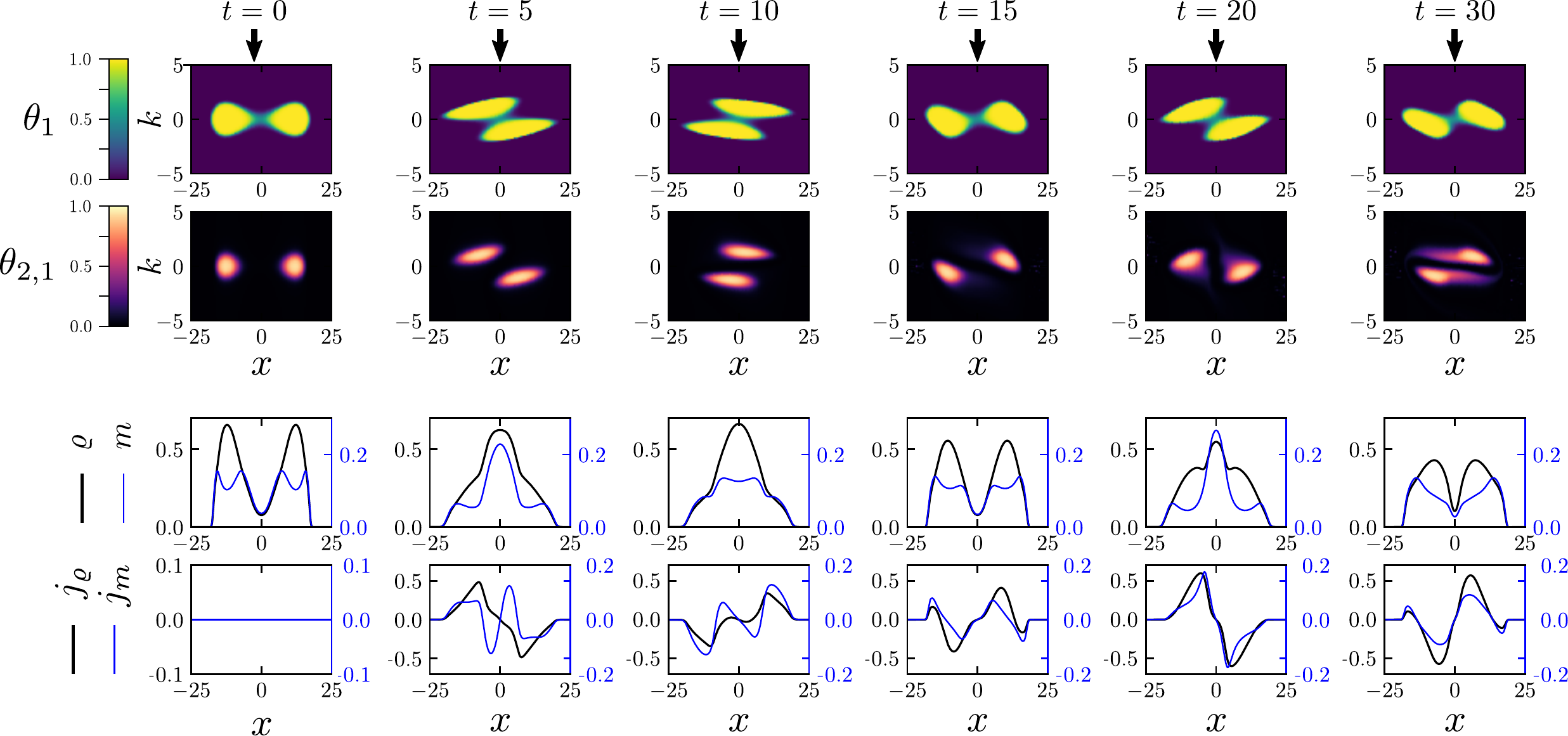}\\
(b)\\[4pt]
\includegraphics[width=0.7\textwidth]{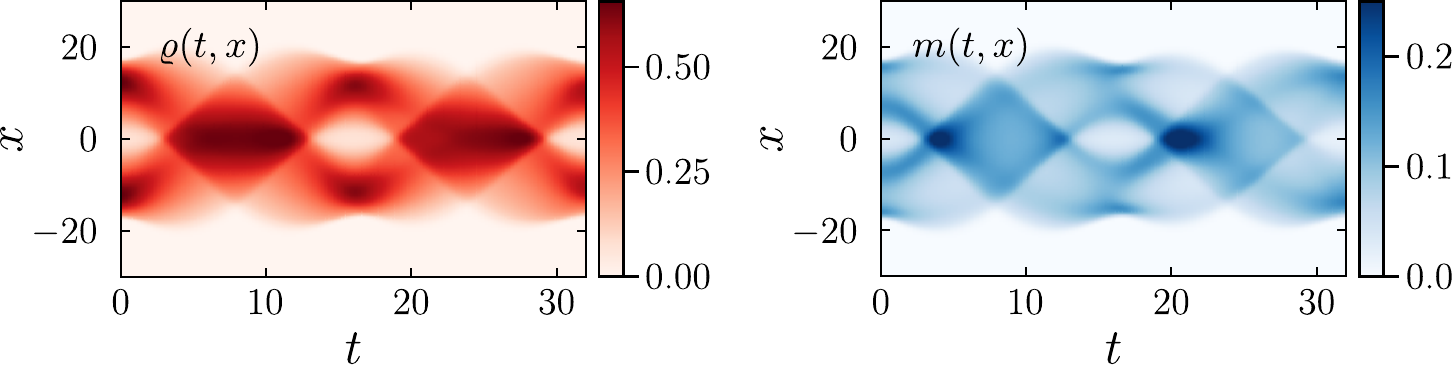}\\
\caption{\label{fig:GHD-QNC}GHD evolution \eqref{ghd} during a quartic-to-quadratic quench: $V(x)=a_4 x^4 -a_2x^2+ \kappa\overset{t=0}{\to} V'(x)=\omega^{'\ 2} x^2$, with $a_4=10^{-4}$, $a_2=3\cdot 10^{-2}$, $\kappa=2.5$ and $\omega'=0.1$. We keep the parameter $\mu=2$, $h=0.5$, $c=1$ fixed and we set the temperature to $T=0.2$. Similarly to Fig.~\ref{fig:GHD-ex1}, the first two rows from the top show the evolution of the filling functions $\theta_1$ and $\theta_{2,1}$ in the rapidity-position plane with times increasing from the leftmost to the rightmost panel. The third and fourth rows show instead the evolution of the particles ({\it thick line}) and magnetization ({\it thin blue line}) density and current as function of the position $x$, with different scales for a better visualization of the curves (left:~particles density; right:~magnetization, {\it in blue}). (b) Colormap of the particles density (left panel, in reds scale) and of the magnetization density (right panel, in blues scale) during the post-quench dynamics.}
\end{figure*}

\section{Zero-Temperature limit}\label{zeroT}

The GHD equations become particularly simple in the limit $T\to 0$, where they allow for further analytic insight, as we discuss in this section. As a preliminary step, we first discuss the $T\to 0$ limit of the model at equilibrium, and review its ground-state phase diagram. 

\subsection{The phase diagram}

We begin by recalling that the TBA equations~\eqref{TBA} admit an alternative, partially decoupled form~\cite{Takahashi1999,Mestyan2019}
\begin{subequations}
\be\begin{split}
\frac{\epsilon_1(k)}{T}=&\frac{k^2-\mu-h}{T} - [r\ast \log(1+ e^{-\epsilon_1/T})](k) \\
&- [s\ast\log(1+e^{\epsilon_{2,1}/T})](k),
\end{split}\ee
\be
\frac{\epsilon_{2,1}(k)}{T}= \left[s\ast \log\left(\frac{1+ e^{\epsilon_{2,2}/T}}{1+e^{-\epsilon_1/T}}\right)\right](k),
\ee
\be\label{tba-dec3}
\frac{\epsilon_{2,n\geq 2}(k)}{T}= \left[s\ast\log\left((1+ e^{\epsilon_{2,n-1}/T})(1+e^{\epsilon_{2,n+1}/T})\right)\right](k),
\ee
\end{subequations}
where
\be
s(k)=\frac{{\rm sech}(\pi k/c)}{2c}, \quad r(k)=[\phi_1\ast s](k).
\ee
These equations must be supplemented by the asymptotic condition
\be
\lim_{n\to\infty} \frac{\epsilon_{2,n}}{T}= \frac{2nh}{T}\,.
\ee
These equations simplify considerably in the limit $T\to 0$. Using that the quasienergies $\epsilon_{2,n}(k)$ are always non-negative, we see that the bound states of spin quasiparticles, corresponding to $n\geq 2$ in the above equations, are exponentially suppressed at low temperatures. Therefore, at $T=0$, we have \cite{Takahashi1999,Batchelor2010,Mestyan2019}
\begin{subequations}\label{tba}
\be\label{tba1}
\epsilon_1(k)=k^2-\mu-h +[\phi_1, \epsilon_{2}]_{\bm 2}(k),
\ee
\be\label{tba2}
\epsilon_{2}(k)=2h +[\phi_1 ,  \epsilon_1]_{\bm 1}(k) -[ \phi_2  , \epsilon_{2}]_{\bm{2}}(k),
\ee
\end{subequations}
where the $[\cdot,\cdot]_{\bm a}$ operation is defined as
\be
[g_1,g_2]_{\bm a}(k) = \int_{-Q_a}^{Q_a} \dd k' \ g_1(k-k') g_2(k'), \quad a=1,2.
\ee
Here, $\vec{Q}=\{ Q_1,Q_2\}$ are cutoffs in the rapidity space for the two quasiparticles species, typically referred to as {\it Fermi points}. They are defined from the condition
\be
\epsilon_a(\pm Q_a)=0, \quad a=1,2.
\ee
Accordingly, the BGT equations~\eqref{BGT} also simplify to
\begin{subequations}\label{bgt}
\be\label{bgt1}
\rho_1(k)=\frac{1}{2\pi}+ [\phi_1,\rho_{2}]_{\bm 2}(k),
\ee
\be\label{bgt2}
\rho_{2}(k)=[\phi_1,\rho_1]_{\bm 1}(k) - [\phi_2,\rho_{2}]_{\bm 2}(k),
\ee
\end{subequations}
while Eqs.~\eqref{BEL}  become
\begin{subequations}\label{bel}
\be\label{v1}
v_1(k) \rho_1(k)= \frac{k}{\pi} +[\phi_1,\rho_{2}v_2]_{\bm 2}(k),
\ee
\be\label{v2}
v_{2}(k) \rho_{2}(k)=[\phi_1,\rho_1 v_1]_{\bm 1}(k) -[\phi_2,\rho_{2} v_{2}]_{\bm 2}(k).
\ee
\end{subequations}
Once again, these equations can be solved by elementary methods. An example of our numerical solution is reported in Fig.~\ref{fig:TBA_ex_zero}. 

\begin{figure}[t!]
\centering
\includegraphics[width=0.35\textwidth]{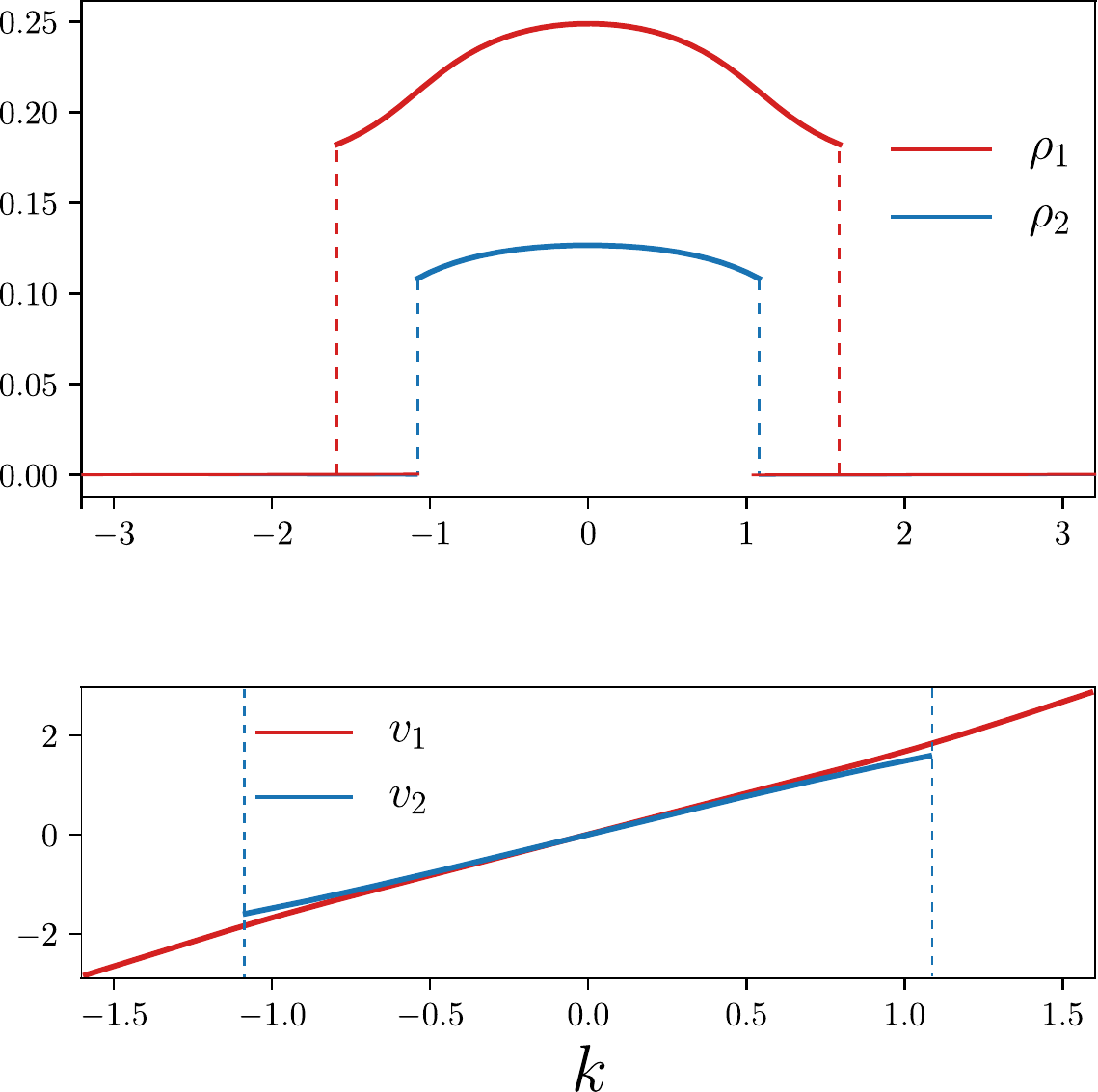}
\caption{\label{fig:TBA_ex_zero}(Top) Root densities $\vec\rho=\{\rho_1,\rho_2\}$ and (bottom) effective velocities $\vec{v}=\{v_1,v_2\}$ of the Yang-Gaudin model at $T=0$. The curves are obtained from the numerical solution of Eqs.~\eqref{bgt} (top) and Eqs.~\eqref{bel} (bottom) with kernel inversion method. We have set $c=1$, $\mu=2$ and $h=0.5$.}
\end{figure}

The above equations also allow us to investigate the phase diagram of the model~\cite{Batchelor2010,He2020}, which we report in Fig.~\ref{fig:phase_diag}. We see the presence of a critical magnetic field $h_c$, which separates a polarized from a partially polarized phase and reads
\be\label{phase_bound} 
\begin{split}
h_c &+ \frac{1}{\pi} \Big[\frac{c}{2}\sqrt{\mu + h_c}\\
& -\left(\mu +h_c+ \frac{c^2}{4}\right) \arctan\left(\frac{\sqrt{\mu + h_c}}{c/2}\right)\Big]=0\,.
\end{split}
\ee
\begin{figure}[t]
\centering
\includegraphics[width=0.35\textwidth]{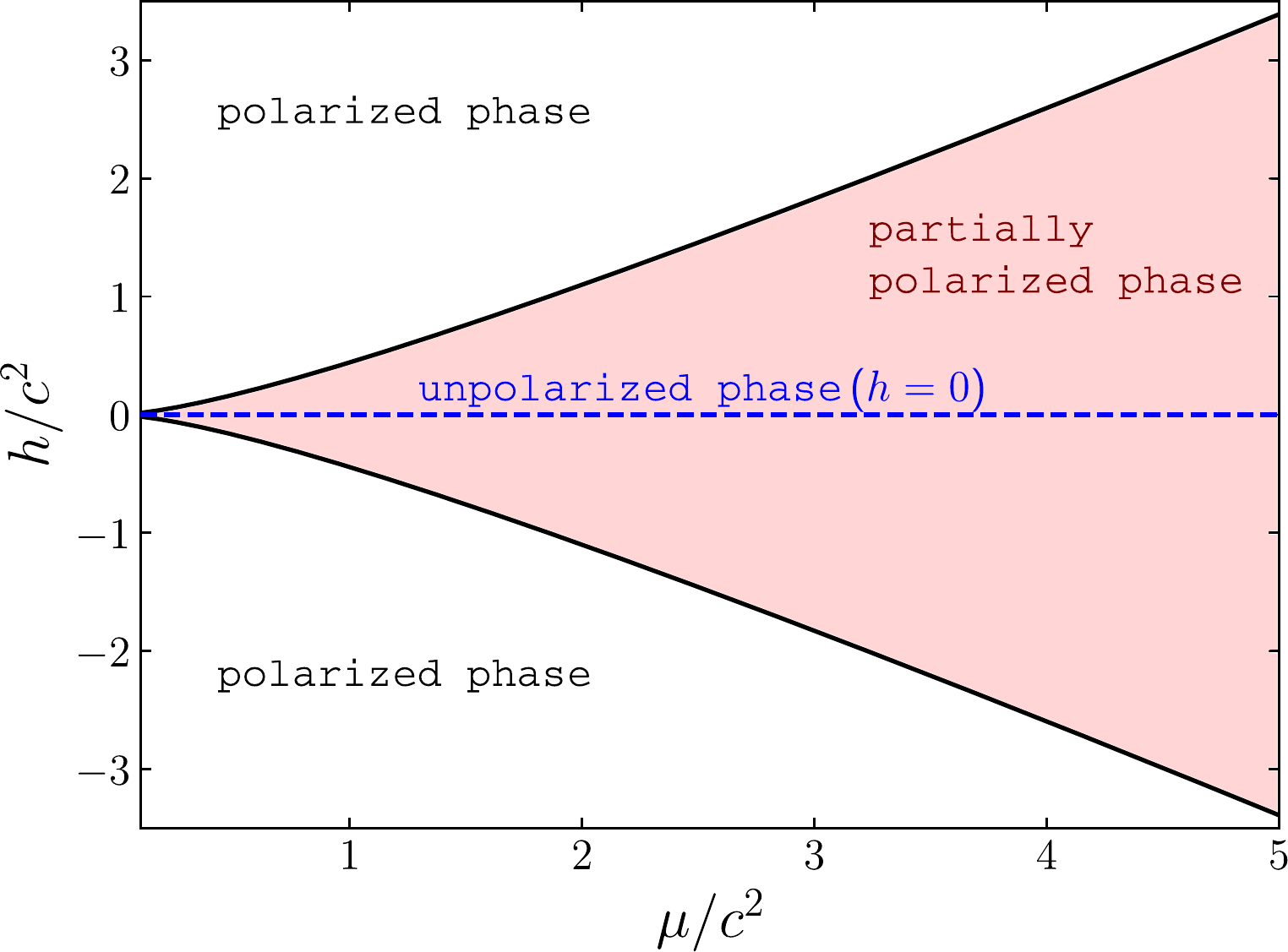}
\caption{\label{fig:phase-diag} Phase diagram for the Yang-Gaudin model at zero temperature. The phase boundary $h=h_c$ (solid line) is obtained from the numerical solution of Eq.~\eqref{phase_bound}.}
\label{fig:phase_diag}
\end{figure}
More precisely, by varying the magnetic field $h$ at fixed $\mu$ and $c$, one finds three distinct phases: \begin{itemize}
\item[(i)]~polarized phase for $h\geq h_c$, which is characterized by a ferromagnetic ground state and by the absence of spin quasiparticles ($Q_2=0$). The first species behaves as a gas of spinless non-interacting particles with Fermi point given by $Q_1=\sqrt{\mu+h}$;
\item[(ii)]~partially polarized phase for $h_c<h<0$, which is characterized by a paramagnetic ground state and by the presence of both the quasiparticles species with Fermi points $\vec{Q}\sim{\cal O}(1)$ extracted from Eqs.~\eqref{tba};
\item[(iii)]~unpolarized phase at $h=0$ where the ground state is a diamagnet and the spinwaves are unbounded, $Q_2\to\infty$, while $Q_1$ is extracted from Eqs.~\eqref{tba}.
\end{itemize}
We briefly comment on the behavior of the Fermi velocities $v^F_a\equiv v_a(Q_a)$ in the limit $h\to 0$, where the spin species becomes unbounded in the rapidity space. From the numerical analysis of Fig.~\ref{fig:vel-sep}, we observe that the Fermi velocity $v^F_2$ monotonically increases for $h< h_c$ and it remains finite approaching the unpolarized phase for $h\to 0$. This feature is reminiscent of the behavior of the Fermi velocity in integrable spin chains at $h=0$, where analytical results for the Fermi velocity are available, see e.g. \cite{Korepin2010}. Importantly, the Fermi velocities of the two species are always separated, which is a crucial feature to observe a dynamical separation of the elementary excitations, see Ref.~\cite{Scopa2021} and Sec.~\ref{SCS}.
%
\begin{figure}[t]
\centering
\includegraphics[width=0.35\textwidth]{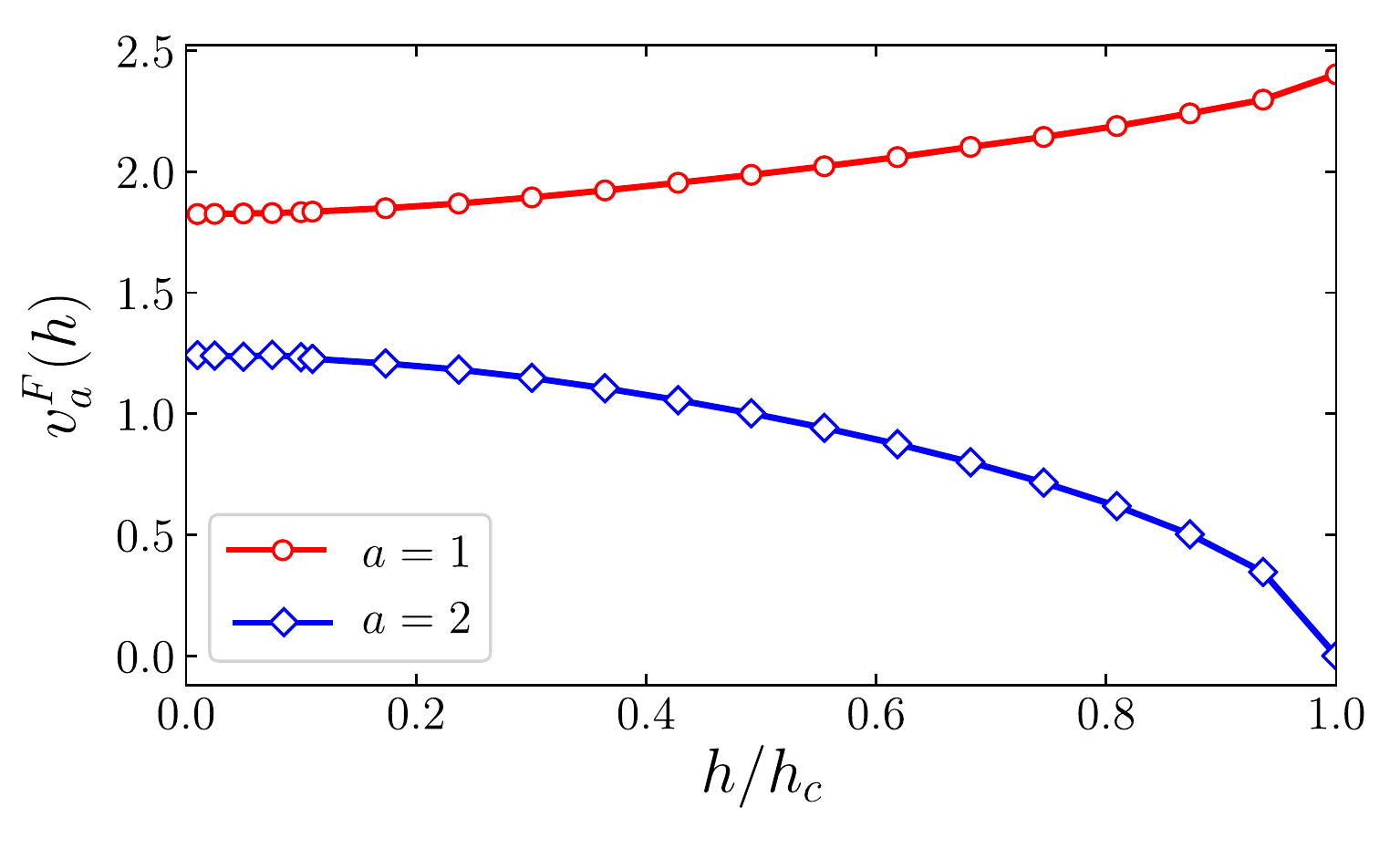}
\caption{\label{fig:vel-sep} Fermi velocities of the Yang-Gaudin model at zero-temperature as function of $h/h_c$. In the figure, we have set $\mu=c=1$.}
\end{figure}
%
\subsection{Zero-entropy GHD\label{sec:zero-entropyGHD}}
We finally return to the inhomogeneous Yang-Gaudin model~\eqref{H} to discuss its generalized hydrodynamic description at zero temperature. As previously argued, under some smoothness assumptions on the potential $U_\sigma(x)$, a local density approximation at mesoscopic scales provides a reliable initial condition for the state at $T=0$, which is the ground state of the model at each spatial point $x$. The latter is suitably described in terms of the local filling functions
\be
\vec\theta(x,k)=\begin{cases} 1, \quad \text{if $k \in [-\vec{Q}(x),\vec{Q}(x)]$}; \\ 0, \quad\text{otherwise}\end{cases}.
\ee
Filling functions of this form correspond to a vanishing Yang-Yang entropy~\cite{Takahashi1999}. Such zero-entropy condition is then preserved at later times by the structure of the GHD equations \eqref{ghd}, as pointed out in Ref.~\cite{Doyon2017}. It follows that the information about the dynamics of $\vec\theta$ can be entirely encoded in a pair of Fermi contours $\vec\Gamma(t)$, keeping track of the evolution of the local Fermi rapidities $\vec{Q}_\alpha(t,x)$ in the rapidity-position plane. More precisely, we can define a pair of split Fermi seas at position $x$ and time $t$ as
\be
\Gamma_a(t,x)\equiv \bigcup_{\alpha=1}^{l_a} \left[{Q}_{a;2\alpha-1}(t,x),{Q}_{a; 2\alpha}(t,x)\right]\,,
\ee
and the Fermi contours
\be
\Gamma_a(t)=\bigcup_x \Gamma_a(t,x),
\ee
where ${Q}_{a;1}(t,x)< {Q}_{a;2}(t,x)< \dots< {Q}_{a;2l_a}(t,x)$ are solutions of the zero-entropy GHD equations \cite{Doyon2017,Ruggiero2020,Scopa2021}:
\be\label{zero-GHD-eqs-gen}
\frac{\dd}{\dd t} \bV x \\[4pt] Q_{a; \alpha}(t,x)\eV =\bV v_{a;\alpha}^F(t,x) \\[4pt] a_a(x,Q_{a;\alpha}(t,x))\eV
\ee
with species index $a=1,2$ and split Fermi sea index $\alpha=1,\dots, 2l_a$. We recall that the effective accelerations $a_a(x,k)$ are those appearing in Eq.~\eqref{acc-gen} while the Fermi velocities $\vec{v}^F_{\alpha}(t,x)\equiv \vec{v}(t,x,\vec{Q}_{\alpha}(t,x))$ with $\vec{v}$ given in Eq.~\eqref{bel}, provided the definition of $[\cdot,\cdot]_{\bm a}$ for the case with split Fermi seas
\be
[g_1,g_2]_{\bm a} =\sum_{\alpha=1}^{l_a} \int_{Q_{a;2\alpha-1}}^{Q_{a; 2\alpha}} \dd k' \ g_1(k-k') g_2(k'), \quad a=1,2.
\ee
In the case of a spatially homogeneous post-quench magnetic field ($g'=0$), the expression for the effective acceleration simplifies and the zero-entropy GHD equations \eqref{zero-GHD-eqs-gen} become
\be\label{zero-GHD-eqs}
\frac{\dd}{\dd t} \bV x \\[4pt] Q_{a; \alpha}(t,x)\eV =\bV v_{a;\alpha}^F(t,x) \\[4pt] -\de_x V'\eV,
\ee
where $V'$ is the post-quench confining potential coupled to the system at $t>0$.\\
\begin{figure}[t!]
\centering
\includegraphics[width=.475\textwidth]{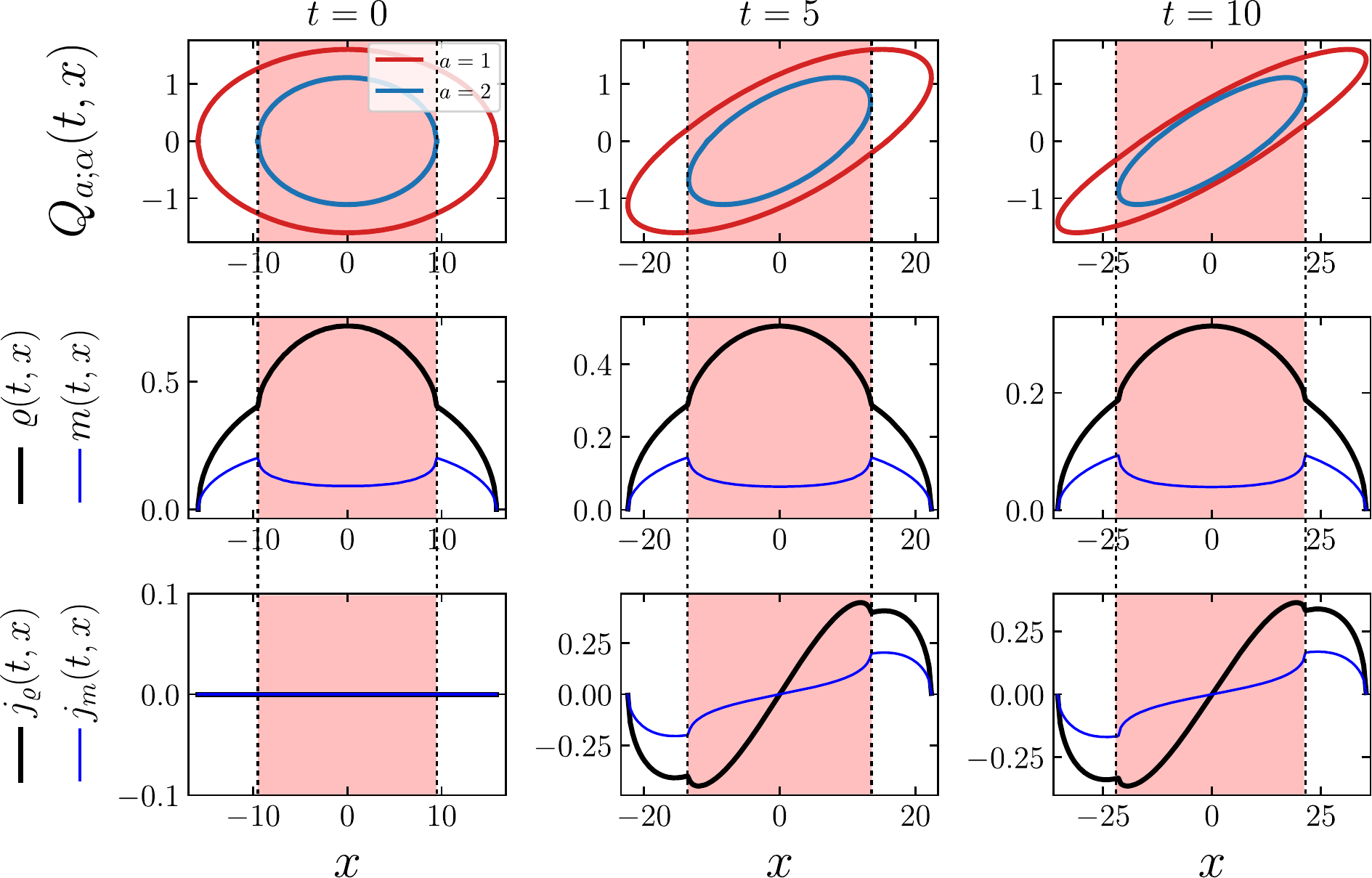}
\caption{\label{fig:zeroT-GHD}Zero-entropy GHD evolution of Yang-Gaudin model suddenly released from a harmonic trap with frequency $\omega=0.1$. We set $c=1$, $\mu=2$ and $h=0.5$. In particular, we show the evolution of the Fermi points $\vec{Q}_{\alpha}(t,x)$ (top row), the particle ({\it thick line}) and magnetization ({\it thin line}) densities (middle row) and currents (bottom row) as function of time, increasing from the left column to the right. The dashed vertical lines mark the position of the transition front $x^\ast(t)$ and the colored area highlight the partially polarized region on each panel.}
\end{figure}

\section{Dynamical polarization and spin-charge separation effects}
\label{sec:zero_t_features}

In this section we focus on two features of the quench dynamics which arise due to the multi-component nature of the elementary excitations, and which are particularly clear to study at zero temperature. They include a dynamical polarization and the spin-charge separation effects. In the latter case, we will also explore its robustness against nonzero temperatures, expanding the results presented in Ref.~\cite{Scopa2021}.

\subsection{Dynamical polarization}\label{sec:dynamical_polarization}
We begin by describing a dynamical polarization process occurring in the gas during the quench dynamics. For simplicity, we consider the case of a harmonic trap $V(x)=\omega^2x^2$ but the following discussion straightforwardly extends to other confining potentials.\\
In this case, the inhomogeneous ground state displays different local phases, determined by the mutual value of the effective chemical potential $\mu-V(x)$ and of the external magnetic field $h$. For sufficiently weak $h\lesssim h_c(\mu,c)$, by moving from the center to the edges of the trap, we can detect a phase transition occurring at positions $\pm x^\ast$ such that $h=h_c(\mu-V(\pm x^\ast),c)$ and separating a partially polarized phase ($|x|\leq x^\ast$) from the fully polarized phases ($|x|>x^\ast$). In general, a similar scenario is found for any trap $V(x)$ that confines the particles in a finite spatial region.\\
By releasing the trap $\omega\overset{t=0}{\to}0$ at fixed value of $\mu$ and $h$, the regions where the system is polarized change dynamically. In particular, the gas remains depolarized in the region $-x^\ast(t)\leq x\leq x^\ast(t)$, and polarized outside of it. The locations of the phase boundary $\pm x^\ast(t)$ during the nonequilibrium dynamics are obtained from the zero-entropy GHD solution as $x^\ast(t)=\max_x[x\in Q_2(x,t)]$.
A similar definition of the phase boundary $x^\ast(t)$ can be derived for more generic quench protocols. In Fig.~\ref{fig:zeroT-GHD}, we show the dynamics of the Fermi points $\vec{Q}_\alpha$ in the rapidity-position plane and the corresponding nonequilibrium evolution of some conserved charges and currents. From the figures one clearly sees a propagating front at position $x^\ast(t)$ that separates the partially polarized regions of the expanding gas (colored areas) from the fully polarized tails. We note that, since the velocity of the first quasiparticle species is larger, the size of the region where the particle density is non-zero but the gas is polarized is effectively growing in time, despite $|x^\ast(t)|$ is increasing.
\\
\begin{figure}[t!]
\centering
\includegraphics[width=0.4\textwidth]{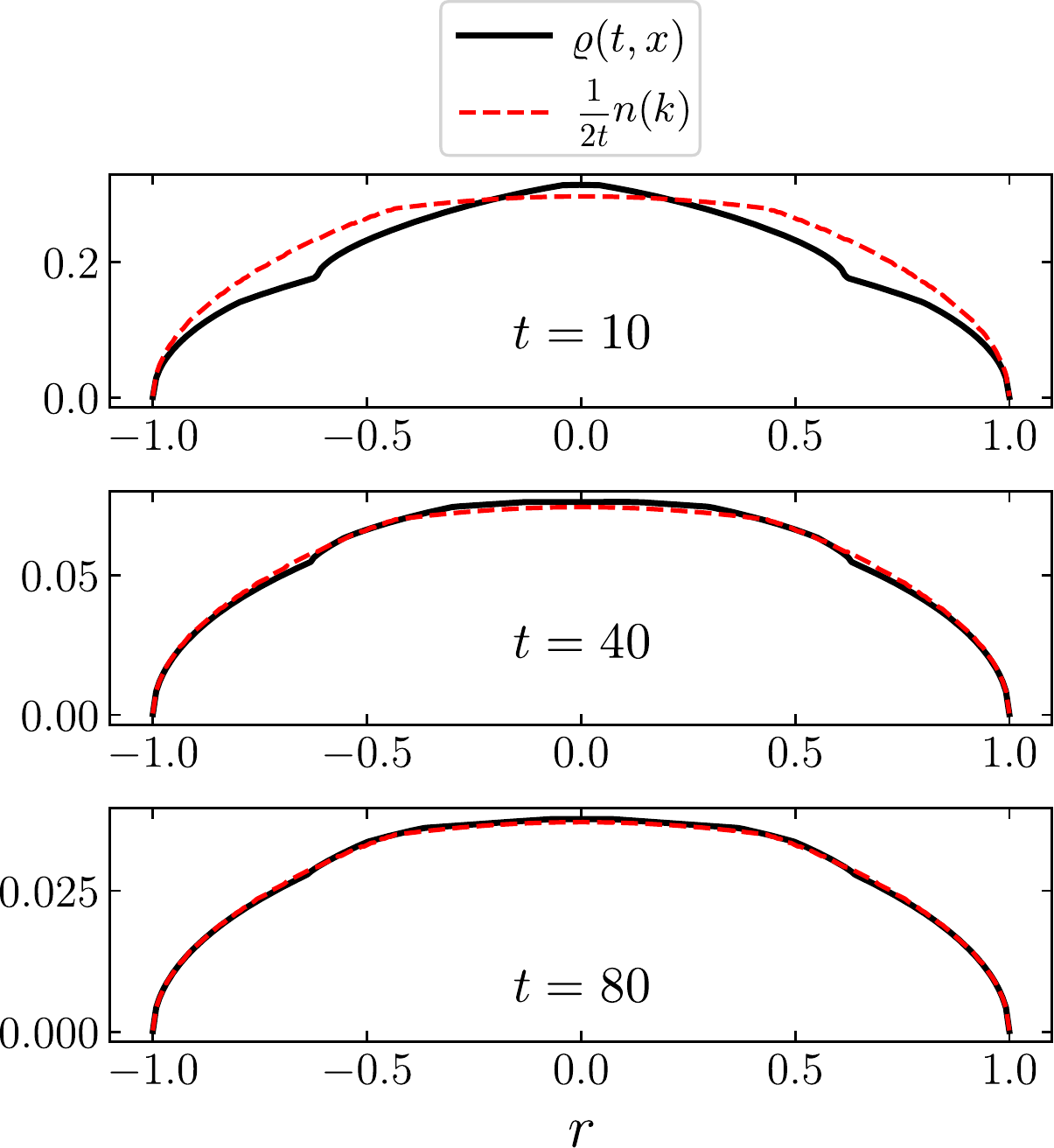}
\caption{\label{fig:Gang}Long-time asymptotic behavior of the particle density ({\it solid line}) during the harmonic trap release of Fig.~\ref{fig:zeroT-GHD} as function of the rescaled position $r\equiv x/R(t)$ where $[-R(t),R(t)]$ is the instantaneous support of the expanding gas. At sufficiently large times, the density profile is indistinguishable from the integrated distribution of rapidities in the initial state $n(k)$, after a proper rescaling by $m/t$ ({\it dashed line}), the mass of the particle $m\equiv 1/2$ in our conventions. The curve $\frac{1}{2t}n(k)$ is shown as function of the rescaled rapidity $r\equiv k/\Lambda$, where $\Lambda$ is the largest value of rapidity in the initial state.}
\end{figure}

As a biproduct, it is also interesting to study the long-time asymptotic behavior of the particle density $\varrho(t,x)$. In polarized regions, the gas is effectively non-interacting, due to its fermionic nature. This suggests that the late-time density profile is determined by the initial distribution of rapidities, in analogy to the what happens in expanding Bose gases, undergoing dynamical fermionization~\cite{Campbell2015}. In fact, based on the zero-temperature GHD semiclassical picture described in the previous sections, and following previous papers~\cite{Campbell2015,Minguzzi2005} 
, one arrives at the prediction
\be\label{eq:dynamical_fermionization}
\varrho(x,t\gg 1) \sim \frac{1}{2t} \int \dd x \ \rho_1(0,x,k) \equiv \frac{1}{2t} n(k)\,,
\ee
where the factor $1/2$ comes from the fact that, within our conventions, the mass of the fermions is $m=1/2$. We have tested the validity of~\eqref{eq:dynamical_fermionization} in Fig.~\ref{fig:Gang}, showing excellent agreement.
\subsection{Spin-charge separation effects}\label{SCS}
Finally, we turn to the analysis of spin-charge separation effects following a trap quench. In our previous paper, Ref.~\cite{Scopa2021}, we analyzed SCS taking place at $T=0$ and $h=0$. In trap release protocols, we showed that SCS was visible in the dynamics of the profiles of particle and magnetization densities, which completely decouple up to perturbatively small corrections. These features were established at $T=0$, and were shown numerically to persist up to small finite temperature $T$. Here, we expand our previous results and study the effects of SCS at finite $h\neq 0$. In this case, we do not expect a separation in the profiles of spin and charge, since the modes which diagonalize the bosonized Hamiltonian mix the corresponding degrees of freedom~\cite{Giamarchi-book}. Rather, we expect the formation of two distinct peaks in both the profiles of magnetization and particle density, which signals the presence of decoupled Luttinger liquids~\cite{Mestyan2019}. In the following, we quantitatively confirm these predictions.\\

We consider the nonequilibrium dynamics generated by a Gaussian potential
\be\label{Gauss-trap}
V(x)=-A\exp\left(-x^2/\varsigma\right)
\ee
with $A,\varsigma$ some tunable parameters that control the amplitude and the width of the pulse perturbation. We then suddenly release the Gaussian confinement at $t=0$ and study the dynamics of the bump in the background at fixed $\mu$ and $h$.
Notice that, with a proper tuning of the background parameters $h<h_c(\mu,c)$, it is possible to initiate and evolve the entire system in local realizations of the partially polarized phase, which is necessary to have both quasiparticles species with non-trivial evolution. The results for the evolution of the Fermi point $\vec{Q}_\alpha(t,x)$ (column a) and of the particle density (column b) are shown in Fig.~\ref{fig:Gauss-zeroT}.\\
\begin{figure}[t]
\centering
(a) \hspace{2cm} (b)\\[3pt]
\includegraphics[width=0.5\textwidth]{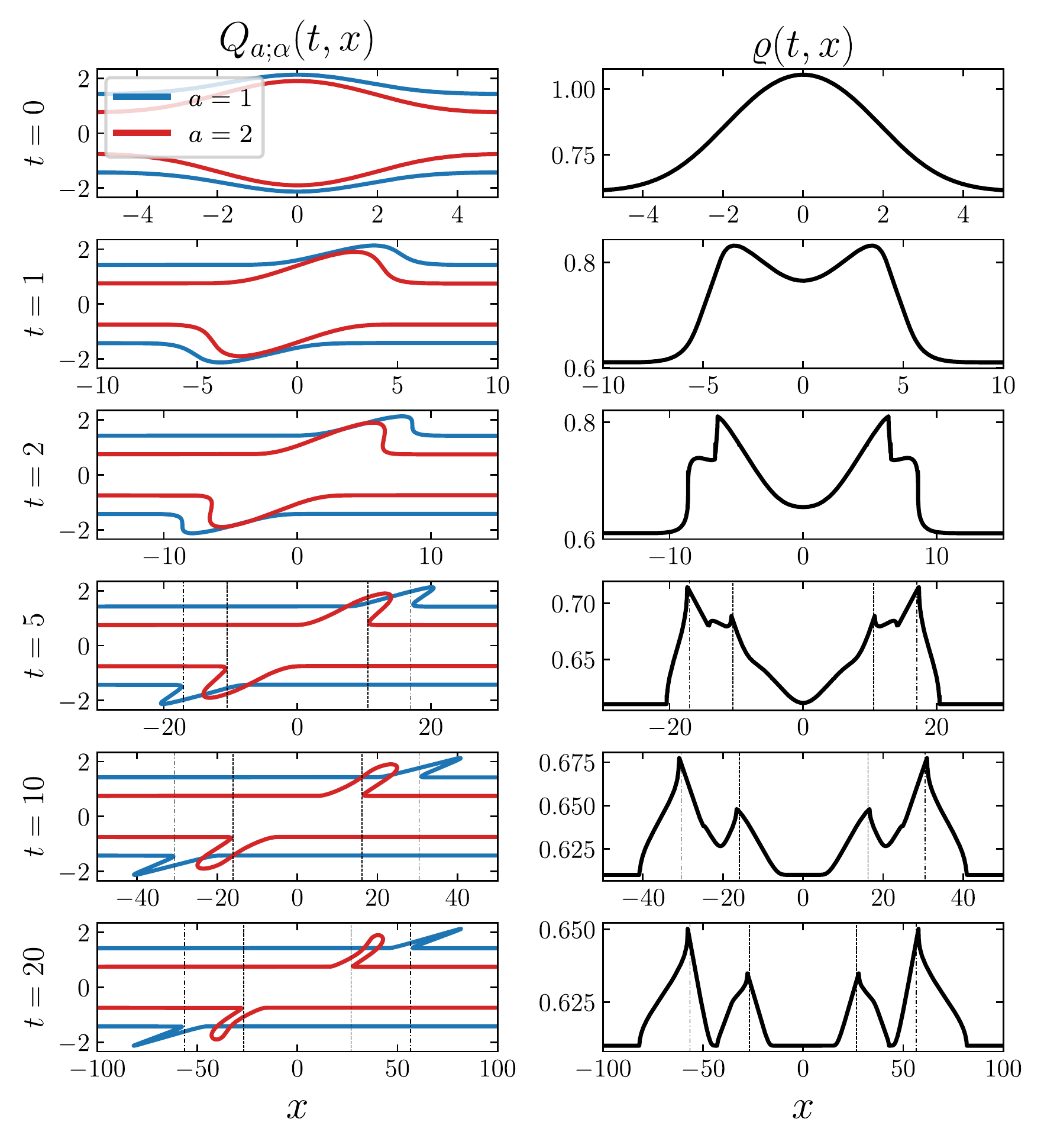}
\caption{\label{fig:Gauss-zeroT}Zero-entropy GHD evolution of the Yang-Gaudin model during the trap release from the Gaussian potential \eqref{Gauss-trap} with $A=2.25$ and $\varsigma=5$. We have set $c=1$, $\mu=1.5$ and $h=0.5$. With this choice of parameters, the system is partially polarized at any spatial position and time during the evolution. We plot the evolution of (a) the Fermi contours and of (b) the corresponding particle density. The dot-dashed (resp.~dashed) vertical axes mark the position of the peaks $|x_1(t)|$ (resp.~$|x_2(t)|$).}
\end{figure}
At $t\leq 0$, we see that the trap \eqref{Gauss-trap} generates a bump perturbation in both the Fermi points of charge (blue curve) and of spin (red curve) species, due to the backreaction effect generated through nested Bethe ansatz. At the physical level, this reflects the spinful nature of the particles in the gas, where the charge and spin degrees of freedom are always mutually activated, even in the presence of selective perturbations (cf Fig.~\ref{fig:zeroT-GHD} where this feature is observed for a harmonic trap). At $t>0$, we track the propagation of the perturbations on the homogeneous background using Eq.~\eqref{zero-GHD-eqs} and we compute the corresponding profiles of the particle density during the post-quench dynamics. We observe that the bump perturbation $Q_{1;\alpha}(t,x)$ moves faster compared to  $Q_{2;\alpha}(t,x)$ and it eventually overturns. This is a direct consequence of the behavior of the Fermi velocities shown in Fig.~\ref{fig:vel-sep}, which is the key to predict the dynamical separation of elementary excitations in the spinful gas. Similar arguments were already put forward long ago based  on the Tomonaga-Luttinger liquid description of the Yang-Gaudin model, see e.g. Ref.~\cite{Giamarchi-book,Recati2003,Recati2003b,Liu2005,Lars2005}. However, GHD, allows us to extend these analyses beyond the low-energy sector.\\

Focusing on the right-propagating quasiparticles, we see that each bump $Q_{a;\alpha}$ gradually deforms in time until generating a shock at $t\approx 2$, after which the species are characterized by a split Fermi sea configuration. Accordingly, at positions $x_a(t)>0$ corresponding to the highest nonsplit Fermi sea of each species, we register peaks in the profiles of the conserved quantities ($a=1$ dot-dashed; $a=2$ dashed vertical axes).  The position of the peaks $x_a(t)$ is simply determined by a ballistic evolution with the Fermi velocity $v^F_a$ of the background. Since $v^F_1>v^F_2$, the innermost peaks are associated to the spinwaves while the outermost refer to the charge excitation. We show this is Fig.~\ref{fig:fronts-gauss} for the pulse perturbation dynamics of Fig.~\ref{fig:Gauss-zeroT}.\\
\begin{figure}[t]
\centering
\includegraphics[width=0.5\textwidth]{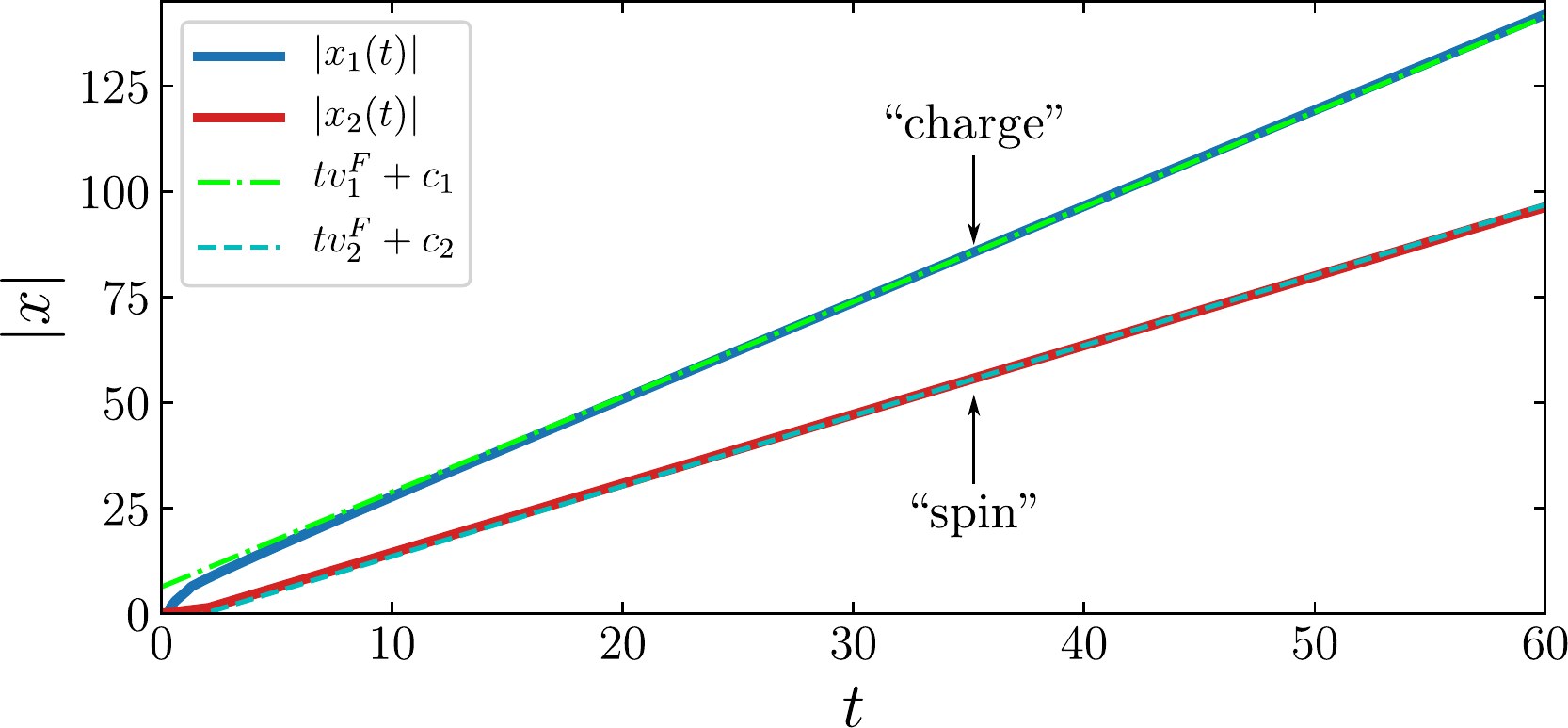}\\
\caption{\label{fig:fronts-gauss} Peaks position $|x_a(t)|$ ({\it solid line}) as function of time for the pulse perturbation dynamics of Fig.~\ref{fig:Gauss-zeroT}. The dot-dashed (dashed) line corresponds to the function $y_a(t) =v_a^F t +c_a$ with slope of the curves given by the Fermi velocities $v_a^F$ in the homogeneous background while the additive constants $c_1\simeq5.3$,  $c_2\simeq6.8$ are extracted from a fit of the data. Notice that away from the line $h=0$ of the phase diagram in Fig.~\ref{fig:phase-diag}, the quasiparticles that split are not the physical charge and spin degrees of freedom of the system but rather a combination of the two.}
\end{figure}
\begin{figure}[t!]
\centering
(a) \hspace{1.5cm} (b)\\
\includegraphics[width=0.8\columnwidth]{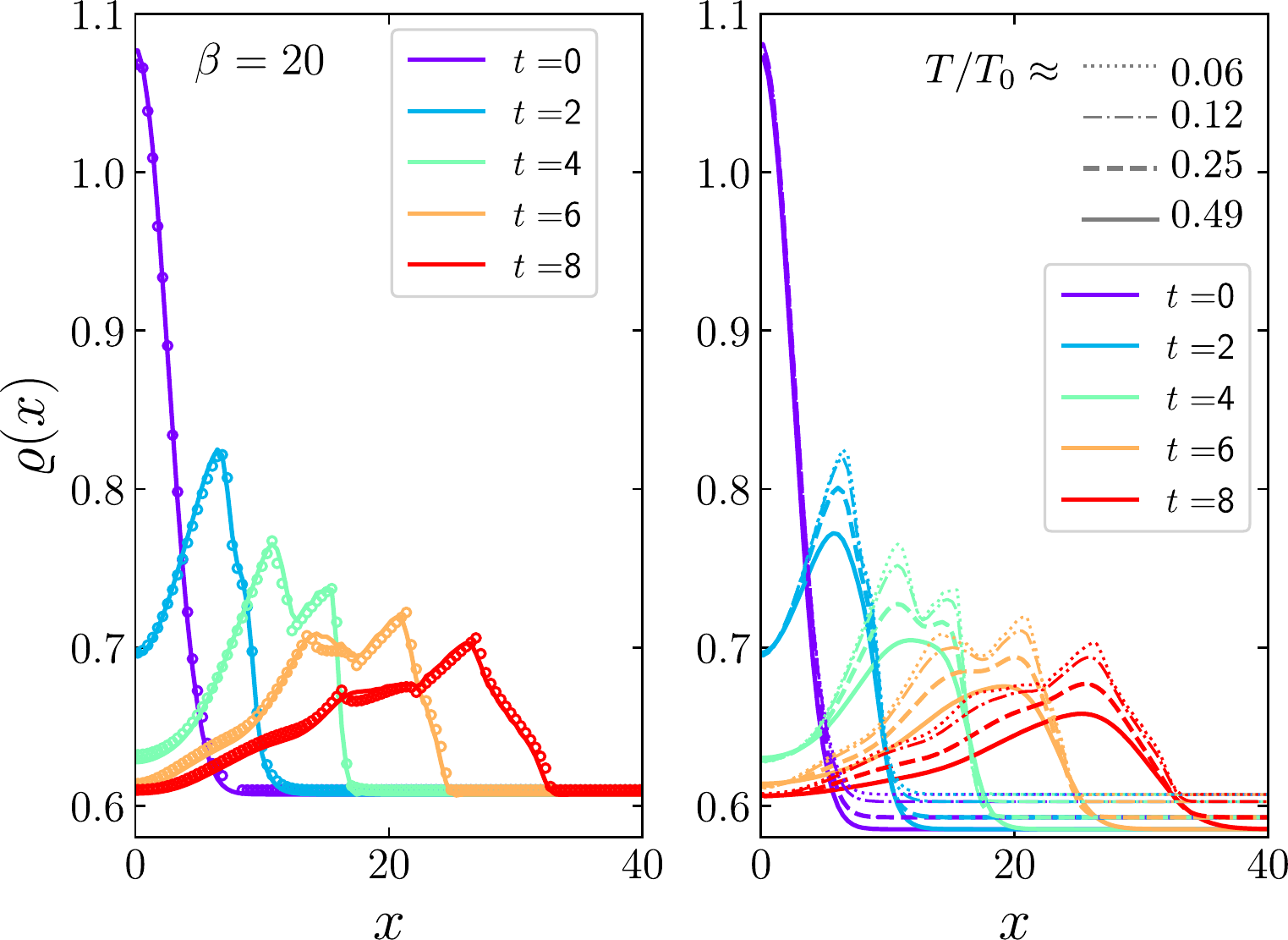}\\
(c)\\
\includegraphics[width=0.95\columnwidth]{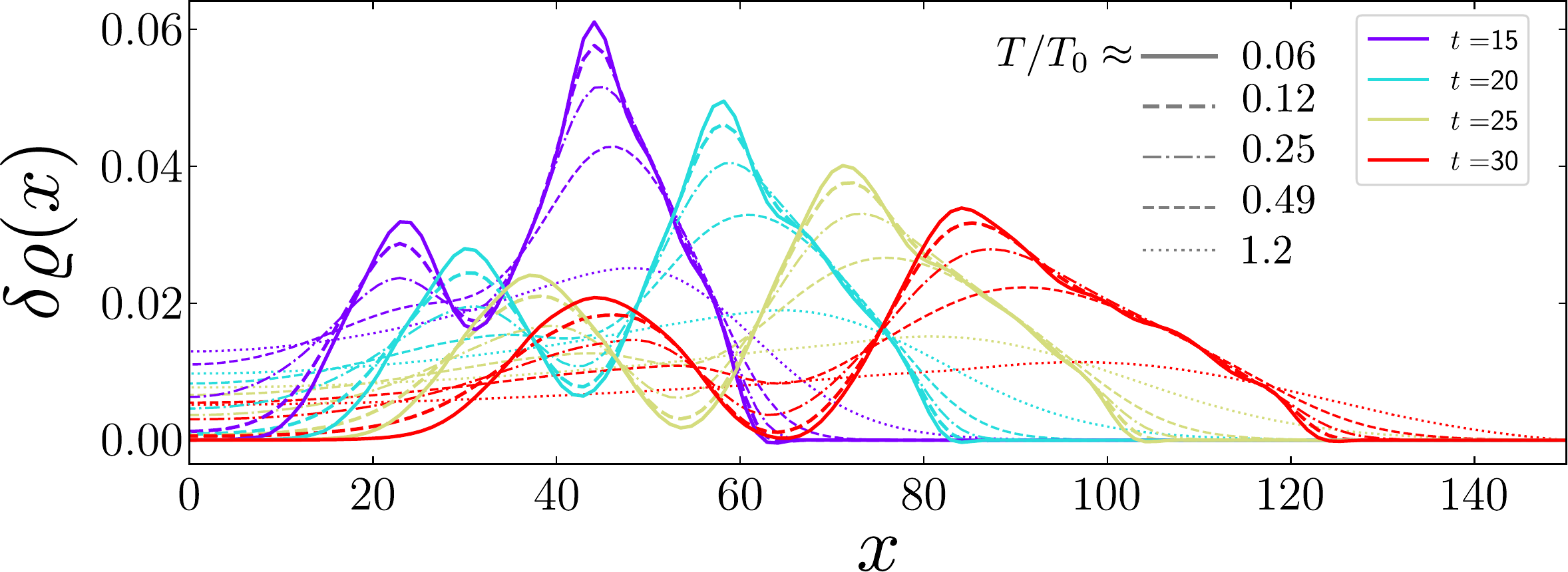}
\caption{\label{fig:Gauss-T}(a) Comparison of the GHD evolution \eqref{ghd} at small but finite temperature $T\equiv\beta^{-1}=0.05$ ({\it solid line}) and of the zero-entropy GHD \eqref{zero-GHD-eqs} ({\it symbols}) for the particle density released from the Gaussian trap \eqref{Gauss-trap} with $A=2.25$ and $\varsigma=10$. We set $c=1$, $\mu=1.5$, and $h=0.5$. (b) Evolution of the particle density at different temperatures $T/T_0$ (different line styles, see plot legend). (c) Long-time evolution of the particle density at different temperatures $T/T_0$. For a better visualization of the curves, we removed the background value $\delta\varrho\equiv\varrho-\varrho_0$ on each of the curves. For our choice of parameters, $T_0\approx 0.81$ and the values of temperature $T/T_0\approx 0.06,0.12,0.25,0.49,1.2$ correspond to inverse temperatures $\beta\equiv T^{-1}=20,10,5,2.5,1$.}
\end{figure}

We finally investigate the effects of a finite temperature $T$  on the pulse perturbation dynamics in Fig.~\ref{fig:Gauss-zeroT}. Intuitively, we expect the separation of elementary excitations to remain clearly visible at small but finite temperature $T\lesssim T_0$ and to be gradually smoothed out by thermal fluctuations after a further increase of temperature $T \gtrsim T_0$. The value of temperature $T_0$ separating these two situations is estimated from the regime of validity of the Tomonaga-Luttinger liquid description of the gas before entering the spin-incoherent regime \cite{Fiete2007,Imambekov2012}, and it reads
\be\label{T0}
T_0\simeq [k_0^2\varrho_0/c]^{-1},
\ee
 with $k_0,\varrho_0$ the Fermi point of first species and the particle density of the partially polarized background, respectively. We test this predictions against GHD and report our results in Fig.~\ref{fig:Gauss-T}, which we now discuss. First, in Fig.~\ref{fig:Gauss-T}(a) we compare the data obtained via zero-entropy GHD \eqref{zero-GHD-eqs} at $T=0$ with those obtained from Eqs.~\eqref{ghd} at small temperature $T\to 0$. The excellent agreement of the profiles constitutes a non-trivial check of our numerical methods. In Fig.~\ref{fig:Gauss-T}(b), we show the results for the nonequilibrium dynamics of the particle density at different temperatures $T/T_0$ while in Fig.~\ref{fig:Gauss-T}(c) we investigate the persistence of density peaks at large times on varying of $T/T_0$. As one can see, our numerical results in Fig.~\ref{fig:Gauss-T}(c) agree with the estimate of the melting temperature in Eq.~\eqref{T0}. 

\section{Summary and conclusions}\label{conclusions}
We investigated the nonequilibrium dynamics generated by the sudden variation of an external inhomogeneous potential in a 1D gas of spin-$\frac{1}{2}$ fermions with repulsive contact interactions, as   described by the inhomogeneous Yang-Gaudin Hamiltonian \eqref{H}.  We have performed a thorough numerical study of the GHD equations, and provided quantitative predictions for different values of the temperature, external magnetic field and chemical potential, highlighting the qualitative features arising due to the multicomponent nature of the elementary excitations. \\

It would be interesting to extend our analysis to more general integrable systems solvable by the nested Bethe ansatz, including multicomponent Bose gases \cite{Fuchs2005,Kollath2008,Kleine2008,Patu2015,Patu2015b,Robinson,Patu2018}, Bose-Fermi mixtures \cite{Patu2019,Wang2020,Patu2022}, or the anyonic interpolation of the bosonic and fermionic Yang-Gaudin models \cite{Patu2019b}. It would also be interesting to better understand how the number of components affects some of the features studied here, for instance in the fermionic Yang-Gaudin model with $K>2$ components. Finally, while we focused on repulsive interactions, several interesting questions pertains to the attractive case. In this context, a natural setting is the one recently considered in Ref.~\cite{Koch2021,Koch2022} for 1D Bose gases, where GHD was exploited to predict the formation of bound states following an adiabatic change of the interactions. We expect that similar protocols in the Yang-Gaudin model could provide valuable insight into its out-of-equilibrium dynamics close to the BEC-BCS transition \cite{BEC-BCS-1,BEC-BCS-2}.\\

{\it Acknowledgements.}~PC and SS acknowledge support from ERC under Consolidator Grant No. 771536 (NEMO). SS is thankful to Alvise Bastianello for useful discussions and collaboration on closely related topics. SS is grateful to the LPENS (Paris) for hospitality at different stages of the development of this work.

\end{document}